\documentclass[a4paper,aps,twocolumn]{revtex4}
\usepackage{amsmath}
\begin{document}
\title{$PT$ Symmetry, Conformal Symmetry, and the Metrication of Electromagnetism}
\author{Philip D. Mannheim}
\affiliation{Department of Physics, University of Connecticut, Storrs, CT 06269, USA.
email: philip.mannheim@uconn.edu}
\date{January 28, 2015}
\begin{abstract}
We present some interesting connections between $PT$ symmetry and conformal symmetry. We use them to develop a metricated theory of electromagnetism in which the electromagnetic field is present in the geometric connection. However, unlike Weyl who first advanced this possibility, we do not take the connection to be real but to instead be $PT$ symmetric, with it being $iA_{\mu}$ rather than $A_{\mu}$ itself that then appears in the connection. With this modification the standard minimal coupling of electromagnetism to fermions is obtained. Through the use of torsion we obtain a metricated theory of electromagnetism that treats its electric and magnetic sectors symmetrically, with a conformal invariant theory of gravity being found to emerge.  An extension to the non-Abelian case is provided.
\end{abstract}
\maketitle

\section{Introduction}

In contemplating a possible unification of gravity with the other fundamental forces it is very appealing  to seek an approach that is intrinsically geometric. Such an approach was pioneered by Weyl who tried to metricate (geometrize) electromagnetism not that long after Einstein first developed a geometric formulation of gravity itself. Weyl's approach involved two key ingredients. The first was a generalization of the Levi-Civita connection $\Lambda^{\lambda}_{\phantom{\alpha}\mu\nu}=(1/2)g^{\lambda\alpha}(\partial_{\mu}g_{\nu\alpha} +\partial_{\nu}g_{\mu\alpha}-\partial_{\alpha}g_{\nu\mu})$ to include an  $A_{\mu}$-dependent geometric Weyl connection $W^{\lambda}_{\phantom{\alpha}\mu\nu}=-g^{\lambda\alpha}(g_{\nu\alpha}A_{\mu} +g_{\mu\alpha}A_{\nu}-g_{\nu\mu}A_{\alpha})$ where $A_{\mu}$ is the electromagnetic vector potential. And the second was the imposition of a real, local scale or conformal transformation that both $g_{\mu\nu}$ and $A_{\mu}$ participate in according to $g_{\mu\nu}(x)\rightarrow e^{2\alpha(x)}g_{\mu\nu}(x)$, $A_{\mu}(x)\rightarrow A_{\mu}(x)+\partial_{\mu}\alpha(x)$, a transformation under which both $\Lambda^{\lambda}_{\phantom{\alpha}\mu\nu}+W^{\lambda}_{\phantom{\alpha}\mu\nu}$ and $\nabla_{\mu}A_{\nu}-\nabla_{\nu}A_{\mu}$ are left invariant. However, as such, the attempt was not successful since in the presence of this Weyl connection the covariant derivative of the metric is non-zero and one has a Weyl geometry rather than a Riemannian one,  with parallel transport being path dependent, and with the state of a system at any given moment being dependent on its prior history. 

Following the subsequent development of quantum mechanics, Weyl's real scale transformation was replaced by a complex gauge transformation that acts on $A_{\mu}$ and electrically charged fields but does not act on $g_{\mu\nu}$ at all, with $A_{\mu}$ being replaced by $iA_{\mu}$ in the coupling to charged fields and with $A_{\mu}$ not appearing  in the geometric connection at all. However, by taking $A_{\mu}$ out of the geometric connection Weyl's attempt to metricate electromagnetism is lost, and one is no longer able to metricate electromagnetism this way.  Also one loses the local scale invariance that Weyl had introduced, a now potentially attractive desideratum for fundamental theory since at the level of the Lagrangian a universe consisting of massless fermions and massless gauge bosons interacting via dimensionless coupling constants is locally conformal invariant.

Since a possible metrication of electromagnetism remains an attractive objective, it is of interest to see if one could revisit Weyl's original program and modify it in some way so that it then would be acceptable. In this paper we present such a possibility. Specifically, we show that, if, in complete parallel to the way one treats the coupling of $A_{\mu}$ to charged fields, one replaces $A_{\mu}$ by $iA_{\mu}$ in the Weyl  geometric connection itself, one is then able to produce a fully acceptable metrication of electromagnetism in which $A_{\mu}$ is then coupled to charged fields in none other than the standard minimal coupling way, with the parallel transport problem even being solved as well, since  parallel transport then turns out to be given just as it is in standard Riemannian geometry. Moreover, this approach generalizes to non-Abelian vector gauge fields as well, and via torsion to axial gauge fields, to thus lead to a metrication of all the fundamental forces.

The key step that is taken in this paper is to focus not on the gauge and metric fields but on fermion fields instead, and in particular to note, that when inserted into the Dirac action Weyl's purely real generalized connection ($A_{\mu}$ being real) turns out to drop out identically and thus not couple to charged fermions at all. Thus Weyl's geometric connection never could have described electromagnetism in the first place. Despite this, we have found that if we  replace $A_{\mu}$ by $iA_{\mu}$ in Weyl's  geometric connection to thereby make it complex, the vector potential then does couple to fermions, and not only does it then do so, it is found to couple just as minimal coupling requires. Since the standard Riemannian Levi-Civita connection  $\Lambda^{\lambda}_{\phantom{\alpha}\mu\nu}$ is based on $\partial_{\mu}$, the prescription is to generalize $\partial_{\mu}$ to $\partial_{\mu}-2iA_{\mu}$ in the geometric connection, rather than to replace $\partial_{\mu}$ by $\partial_{\mu}-2A_{\mu}$ in the geometric connection (the original Weyl prescription). (Under the discrete antilinear $PT$ transformation that we consider below it is $iA_{\mu}$ that transforms the same way as $\partial_{\mu}$ and not $A_{\mu}$ itself.) With this now complex geometric connection we obtain a completely dual description of electromagnetism, either via coupling to fermions in the standard local minimal way or via an $iA_{\mu}$-dependent geometric connection, with the $\int d^4x(-g)^{1/2}i\bar{\psi}\gamma^{a}V^{\mu}_a(\partial_{\mu}+\Sigma_{bc}\omega^{bc}_{\mu}-iA_{\mu})\psi$  action that ensues being the selfsame one in the two cases, and with electromagnetism being metricated.

Now, with such an $iA_{\mu}$-dependent geometric connection the immediate concern is that $g_{\mu\nu}$ would then couple through a then generalized, and even complex, Riemann tensor and  lead to a gravity theory that does not look anything like the gravity that is observed.  However, because of Weyl's very same conformal invariance this does not in fact occur. Specifically, for the standard minimal coupling of fermions to $A_{\mu}$ to be able to possess Weyl's  local conformal invariance, $A_{\mu}$ would have to have conformal weight zero and not transform under a conformal transformation at all, viz. $A_{\mu}\rightarrow A_{\mu}$, (just as $g_{\mu\nu}$ does not transform under an electromagnetic gauge transformation). Because of this, the only geometric action one could write down that would be locally conformal invariant would be the one based on the square (viz. $C_{\mu\nu\sigma\tau}C^{\mu\nu\sigma\tau}$) of the Weyl tensor $C_{\mu\nu\sigma\tau}$ (cf. (\ref{47}) and (\ref{48}) below) as constructed via the Levi-Civita connection alone, with a generalized Weyl tensor built out of the Levi-Civita plus $iA_{\mu}$-dependent Weyl connection not being locally conformal invariant for an $A_{\mu}$ that does not transform under a conformal transformation. Now such a  generalized Weyl-connection-dependent Weyl tensor would have been locally conformal invariant had $A_{\mu}$ transformed non-trivially ($A_{\mu}\rightarrow A_{\mu}+\partial_{\mu}\alpha(x)$) under a conformal transformation just as Weyl had originally proposed. However, with an $A_{\mu}$ that does not transform at all under a local conformal transformation, this very same conformal invariance then forces the geometry to depend on the Levi-Civita connection alone, with the Weyl connection contributing solely to and being buried in the coupling of $A_{\mu}$ to the fermions. With the pure metric sector of the theory only depending on the Levi-Civita connection, parallel transport is thus strictly Riemannian, and thus through the replacement of $A_{\mu}$ by $iA_{\mu}$ in the Weyl connection we convert Weyl geometry into Riemannian geometry. 

Thus by making two changes in Weyl's approach, namely by replacing $A_{\mu}$ by $iA_{\mu}$ in the Weyl geometric connection and by giving $A_{\mu}$ the zero conformal weight that the Dirac action requires, one can then construct a metrication of electromagnetism. Moreover, the extension to the strong and weak gauge theories is immediate, since if one also gives the non-Abelian gauge fields  conformal weight zero and couples them to the geometry via an analogous non-Abelian $iA_{\mu}$-type geometric connection, they also do not couple in the geometric $C_{\mu\nu\sigma\tau}C^{\mu\nu\sigma\tau}$ but only in non-Abelian generalizations of $F_{\mu\nu}F^{\mu\nu}$. Since the strong, electromagnetic and weak interactions are based on the non-Abelian $SU(3)\times SU(2)\times U(1)$ local gauge theory, our approach thus permits a metrication of all the fundamental forces, with their chiral aspects being accommodated through the introduction a further geometric connection, namely one with torsion.

Now the reader might be concerned that our results are somewhat restrictive since they appear to require the a priori imposition of local conformal invariance. However, this turns out not to be the case, since one can obtain our results via a completely different procedure. Specifically, if one starts with the Dirac action for a fermion coupled to the geometry via both an $iA_{\mu}$-dependent geometric connection and a standard purely Riemannian Levi-Civita-based spin connection, then, as we discuss in detail below, on doing a path integration on the fermions (equivalent to a fermion one loop Feynman diagram) one obtains an effective action for gravity and electromagnetism that is precisely of the $C_{\mu\nu\sigma\tau}C^{\mu\nu\sigma\tau}$ plus $F_{\mu\nu}F^{\mu\nu}$ form, where the $C_{\mu\nu\sigma\tau}C^{\mu\nu\sigma\tau}$ term is based on the standard Levi-Civita connection alone. With the generalized Weyl connection only appearing as the $F_{\mu\nu}F^{\mu\nu}$ term, and with the $C_{\mu\nu\sigma\tau}C^{\mu\nu\sigma\tau}$ term being based on the standard Levi-Civita connection alone, the geometry is then strictly Riemannian and the $iA_{\mu}$ dependence does not appear in the coupling of the metric to the geometry. Thus the path integration on the fermions serves to produce an effective action for gravity and electromagnetism in which  the Levi-Civita and Weyl connections are completely decoupled. Moreover this decoupling persists even if we extend the theory to non-Abelian $A_{\mu}^i$ and even if we add in torsion as well, and even if we generalize torsion to the non-Abelian case. That this decoupling occurs is because the Dirac action for a fermion coupled to the various Weyl, torsion, and Levi-Civita connections turns out to be locally conformally invariant (up to fermion mass terms), so that a path integration over the fermions will necessarily produce an effective action for the various $g_{\mu\nu}$ and $A_{\mu}$ fields whose leading term is  locally conformal invariant too (the effect of mass is non-leading since the mass term is a soft operator). The fermion path integration thus does the separation of the $g_{\mu\nu}$ and $A_{\mu}$ sectors for us without our needing to impose it in advance.

To underscore the point we note that had we started with a completely conventional Dirac action in which the fermion is coupled to the geometry through a Levi-Civita-based spin connection and coupled to $A_{\mu}$ through conventional minimal coupling (viz. (\ref{33}) below), fermion path integration would generate an effective action containing a purely Riemannian geometry in the $g_{\mu\nu}$ sector and a purely conventional $F_{\mu\nu}F^{\mu\nu}$ term in the $A_{\mu}$ sector (viz. (\ref{43})). We would not at all expect to get an effective action that would involve Weyl geometry, and of course we do not. With the metrication that we present here leading to a complete duality between minimal coupling and the $iA_{\mu}$-based Weyl connection  approach, the Weyl connection approach must generate the selfsame Dirac action, and thus it too must lead to an effective action in which there is a complete separation between the gravity and electromagnetic sectors, with the gravity sector being based on the Levi-Civita connection alone. 

To develop the results presented in this paper we need to explore the interplay of geometric connections with $PT$ symmetry, $CPT$ symmetry, and conformal symmetry. We present the various geometric connections of interest to us in Sec. II, and in Sec. III we present the various $PT$, $CPT$, conformal and Lorentz symmetry aspects  of interest to us here. In Sec. IV we discuss metrication associated with torsion, and in Sec. V we discuss metrication associated with the Weyl connection. Finally, we comment on the fact that our approach leads us to conformal gravity rather than to standard Newton-Einstein gravity. Reviews of torsion may be found in \cite{Hehl1976,Shapiro2002,Hammond2002}, and recent reviews of Weyl geometry may be found in \cite{Scholz2011} and \cite{Yang2014}. A review of $PT$ symmetry ($P$ is parity, $T$ is time reversal) may be found in \cite{Bender2007}. Some recent discussion of conformal gravity may be found in \cite{Mannheim2006,Mannheim2011a,Mannheim2012a} and \cite{tHooft2010a,tHooft2010b,tHooft2011,tHooft2014}.

\section{The Various  Spacetime Connections and the Dirac Action}

\subsection{The Spacetime Connections}

In order to construct covariant derivatives in any curvature-based theory of gravity one must introduce a three-index connection $\Gamma^{\lambda}_{\phantom{\alpha}\mu\nu}$, with the only requirement on it being that  it transform under a coordinate transformation $x^{\mu}\rightarrow x^{\prime \mu}$ as 
\begin{eqnarray}
\Gamma^{\prime\lambda}_{\phantom{\alpha}\mu\nu}(x^{\prime})=
\frac{dx^{\prime\lambda}}{dx^{\alpha}}
\frac{dx^{\beta}}{dx^{\prime \mu}}
\frac{dx^{\gamma}}{dx^{\prime \nu}}
\Gamma^{\alpha}_{\phantom{\alpha}\beta\gamma}(x)+
\frac{d^2x^{\rho}}{dx^{\prime\mu}dx^{\prime \nu}}\frac{dx^{\prime\lambda}}{dx^{\rho}}.
\label{1}
\end{eqnarray}
With this condition covariant derivatives such as
\begin{eqnarray}
\nabla_{\mu}g^{\lambda\nu}=\partial_{\mu}g^{\lambda\nu}+\Gamma^{\lambda}_{\phantom{\alpha}\alpha\mu}g^{\alpha\nu}+\Gamma^{\nu}_{\phantom{\alpha}\alpha\mu}g^{\lambda\alpha}
\label{2}
\end{eqnarray}
transform as true general coordinate tensors, i.e. as
\begin{eqnarray}
\nabla^{\prime}_{\mu}g^{\lambda\nu}(x^{\prime})=
\frac{dx^{\prime\lambda}}{dx^{\alpha}}
\frac{dx^{\prime \nu}}{dx^{ \beta}}
\frac{dx^{\gamma}}{dx^{\prime \mu}}
\nabla_{\gamma}g^{\alpha\beta}(x).
\label{3}
\end{eqnarray}
Moreover, given only that the connection transforms as in Eq. (\ref{1}), the four-index object
\begin{eqnarray}
R^{\lambda}_{\phantom{\rho}\mu\nu\kappa}
=\partial_{\kappa}\Gamma^{\lambda}_{\phantom{\alpha}\mu\nu}-\partial_{\nu}\Gamma^{\lambda}_{\phantom{\alpha}\mu\kappa}
+\Gamma^{\eta}_{\phantom{\alpha}\mu\nu}\Gamma^{\lambda}_{\phantom{\alpha}\eta\kappa}
-\Gamma^{\eta}_{\phantom{\alpha}\mu\kappa}\Gamma^{\lambda}_{\phantom{\alpha}\eta\nu}
\label{4}
\end{eqnarray}
transforms as a true rank four tensor and is known as the Riemann curvature tensor.

For pure Riemannian geometry the connection is given by the Levi-Civita connection 
\begin{eqnarray}
\Lambda^{\lambda}_{\phantom{\alpha}\mu\nu}=\frac{1}{2}g^{\lambda\alpha}(\partial_{\mu}g_{\nu\alpha} +\partial_{\nu}g_{\mu\alpha}-\partial_{\alpha}g_{\nu\mu}),
\label{5}
\end{eqnarray}
and with it the metric obeys the metricity (or metric compatible) condition $\nabla_{\mu}g^{\lambda\nu}=0$.

However, one is free to add on to $\Lambda^{\lambda}_{\phantom{\alpha}\mu\nu}$ any additional rank three tensor  $\delta{\Gamma}^{\lambda}_{\phantom{\alpha}\mu\nu}$ since $\tilde{\Gamma}^{\lambda}_{\phantom{\alpha}\mu\nu}=\Lambda^{\lambda}_{\phantom{\alpha}\mu\nu}+\delta{\Gamma}^{\lambda}_{\phantom{\alpha}\mu\nu}$ will still obey (\ref{1}) if $\delta{\Gamma}^{\lambda}_{\phantom{\alpha}\mu\nu}$ is itself a tensor. In terms of $\tilde{\Gamma}^{\lambda}_{\phantom{\alpha}\mu\nu}$ one defines covariant derivatives such as 
\begin{eqnarray}
\tilde{\nabla}_{\mu}g^{\lambda\nu}=\partial_{\mu}g^{\lambda\nu}+\tilde{\Gamma}^{\lambda}_{\phantom{\alpha}\alpha\mu}g^{\alpha\nu}+\tilde{\Gamma}^{\nu}_{\phantom{\alpha}\alpha\mu}g^{\lambda\alpha},
\label{6}
\end{eqnarray}
and whether or not the metric obeys the generalized metricity condition $\tilde{\nabla}_{\mu}g^{\lambda\nu}=0$ depends on the choice of $\delta{\Gamma}^{\lambda}_{\phantom{\alpha}\mu\nu}$. Additionally, the four index object 
\begin{eqnarray}
\tilde{R}^{\lambda}_{\phantom{\rho}\mu\nu\kappa}
=\partial_{\kappa}\tilde{\Gamma}^{\lambda}_{\phantom{\alpha}\mu\nu}-\partial_{\nu}\tilde{\Gamma}^{\lambda}_{\phantom{\alpha}\mu\kappa}
+\tilde{\Gamma}^{\eta}_{\phantom{\alpha}\mu\nu}\tilde{\Gamma}^{\lambda}_{\phantom{\alpha}\eta\kappa}
-\tilde{\Gamma}^{\eta}_{\phantom{\alpha}\mu\kappa}\tilde{\Gamma}^{\lambda}_{\phantom{\alpha}\eta\nu}
\label{7}
\end{eqnarray}
is also a true tensor. In terms of the Levi-Civita-based derivative $\nabla_{\mu}$ the generalized $\tilde{R}^{\lambda}_{\phantom{\rho}\mu\nu\kappa}$ can be rewritten as 
\begin{eqnarray}
\tilde{R}^{\lambda}_{\phantom{\rho}\mu\nu\kappa}&=&R^{\lambda}_{\phantom{\rho}\mu\nu\kappa}
+\nabla_{\kappa}\delta{\Gamma}^{\lambda}_{\phantom{\rho}\mu\nu}-\nabla_{\nu}\delta{\Gamma}^{\lambda}_{\phantom{\rho}\mu\kappa}
\nonumber\\
&+&\delta{\Gamma}^{\eta}_{\phantom{\alpha}\mu\nu}\delta{\Gamma}^{\lambda}_{\phantom{\alpha}\eta\kappa}
-\delta{\Gamma}^{\eta}_{\phantom{\alpha}\mu\kappa}\delta{\Gamma}^{\lambda}_{\phantom{\alpha}\eta\nu},
\label{8}
\end{eqnarray}
a form which follows since $\delta{\Gamma}^{\lambda}_{\phantom{\alpha}\mu\nu}$ is a true tensor.

Each different choice of $\delta{\Gamma}^{\lambda}_{\phantom{\alpha}\mu\nu}$ defines its own geometry, each with its own $\tilde{R}^{\lambda}_{\phantom{\rho}\mu\nu\kappa}$. Our interest here is two particular connections: the previously introduced Weyl connection 
\begin{eqnarray}
W^{\lambda}_{\phantom{\alpha}\mu\nu}=-g^{\lambda\alpha}(g_{\nu\alpha}A_{\mu} +g_{\mu\alpha}A_{\nu}-g_{\nu\mu}A_{\alpha}),
\label{9}
\end{eqnarray}
and the contorsion connection
\begin{eqnarray}
K^{\lambda}_{\phantom{\alpha}\mu\nu}=\frac{1}{2}g^{\lambda\alpha}(Q_{\mu\nu\alpha}+Q_{\nu\mu\alpha}-Q_{\alpha\nu\mu}),
\label{10}
\end{eqnarray}
where 
\begin{eqnarray}
Q^{\lambda}_{\phantom{\alpha}\mu\nu}=\Gamma^{\lambda}_{\phantom{\alpha}\mu\nu}-\Gamma^{\lambda}_{\phantom{\alpha}\nu\mu}
\label{11}
\end{eqnarray}
is the Cartan torsion tensor associated with a connection that has an  antisymmetric part. With the Weyl connection being symmetric on its two lower indices and the contorsion connection being antisymmetric on them, we can anticipate that these two connections will respectively have some relation to vector and axial vector fields.

Of the two connections the metric obeys a metricity condition when $\delta{\Gamma}^{\lambda}_{\phantom{\alpha}\mu\nu}=K^{\lambda}_{\phantom{\alpha}\mu\nu}$. However it does not do so when $\delta{\Gamma}^{\lambda}_{\phantom{\alpha}\mu\nu}=W^{\lambda}_{\phantom{\alpha}\mu\nu}$, since for it one has $\tilde{\nabla}_{\sigma}g^{\mu\nu}=-2g^{\mu\nu}A_{\sigma}$. While this is actually a quite intriguing relation since it is left invariant  under $g_{\mu\nu}(x)\rightarrow e^{2\alpha(x)}g_{\mu\nu}(x)$, $A_{\mu}(x)\rightarrow A_{\mu}(x)+\partial_{\mu}\alpha(x)$, it nonetheless leads to a path dependence to parallel transport, thereby rendering Weyl geometry untenable as is. 

Nothing that we know of requires us to consider either of these two choices for $\delta{\Gamma}^{\lambda}_{\phantom{\alpha}\mu\nu}$, and nothing would appear to go wrong if  they are not considered. However, they do have certain advantages. Use of the torsion connection provides insights into spin and axial gauge symmetry, and use of the Weyl connection provides insights into vector gauge invariance and conformal invariance. Recently, we have shown \cite{Fabbri2014,Mannheim2014a,Mannheim2014b} that torsion provides insights into both gravitation and electromagnetism. And in this paper we show that these developments are interrelated with $PT$ symmetry  and Weyl geometry in a way that will enable us to both metricate electromagnetism and convert Weyl geometry into standard Riemannian geometry, and thereby dispose of its parallel transport problem. 

\subsection{The Spin Connection}

While one uses the connection $\Gamma^{\lambda}_{\phantom{\alpha}\mu\nu}$ to implement local translation invariance, to implement local Lorentz invariance one introduces a set of vierbeins $V^{a}_{\mu}$ where the coordinate $a$ refers to a fixed, special-relativistic reference coordinate system with metric $\eta_{ab}$, with the Riemannian metric then being writable as $g_{\mu\nu}=\eta_{ab}V^{a}_{\mu}V^{b}_{\nu}$. With the vierbein carrying a fixed basis index its covariant derivatives are not given by $\Gamma^{\lambda}_{\phantom{\alpha}\mu\nu}$ alone. Rather, one introduces a second connection known as the spin connection $\Omega_{\mu}^{ab}$, with it being the derivative
\begin{eqnarray}
D_{\mu}V^{a\lambda}=\partial_{\mu}V^{a\lambda}+\Lambda^{\lambda}_{\phantom{\alpha}\nu\mu}V^{a \nu}+\Omega_{\mu}^{ab}V^{\lambda}_{b}
\label{12}
\end{eqnarray}
that will transform as a tensor under both local translations and local Lorentz transformations provided the spin connection transforms as
\begin{eqnarray}
\Omega_{\mu}^{\prime ab}=
\Lambda^a_{\phantom{a}c}(x)\Lambda^b_{\phantom{b}d}(x)\Omega_{\mu}^{cd}
-\Lambda^{bc}(x)\partial_{\mu}\Lambda^a_{\phantom{a}c}(x)
\label{13}
\end{eqnarray}
under $V^a_{\mu}(x^{\lambda})\rightarrow \Lambda^a_{\phantom{a}c}(x)V^c_{\mu}(\Lambda^{\lambda}_{\phantom{\lambda}\tau}x^{\tau})$. For a standard Riemannian geometry the spin connection is given by 
\begin{eqnarray}
-\omega_{\mu}^{ab}=V^b_{\nu}\partial_{\mu}V^{a\nu}+V^b_{\lambda}\Lambda^{\lambda}_{\phantom{\lambda}\nu\mu}V^{a\nu}, 
\label{14}
\end{eqnarray}
and with this connection the vierbein obeys metricity in the form $D_{\mu}V^{a\lambda}=0$. Finally, when one uses the generalized connection $\tilde{\Gamma}^{\lambda}_{\phantom{\alpha}\mu\nu}=\Lambda^{\lambda}_{\phantom{\alpha}\mu\nu}+\delta{\Gamma}^{\lambda}_{\phantom{\alpha}\mu\nu}$, one must use the generalized spin connection $\tilde{\omega}_{\mu}^{ab}=\omega_{\mu}^{ab}+\delta{\omega}_{\mu}^{ab}$, where 
\begin{eqnarray}
-\tilde{\omega}_{\mu}^{ab}=-\omega_{\mu}^{ab}+V^{b}_{\lambda}\delta{\Gamma}^{\lambda}_{\phantom{\alpha}\nu\mu}V^{a \nu},
\label{15}
\end{eqnarray}
with $\tilde{\omega}_{\mu}^{ab}$ obeying (\ref{13}) if $\tilde{\Gamma}^{\lambda}_{\phantom{\alpha}\mu\nu}$ obeys (\ref{1}). Given the generalized spin connection the metric will only obey the generalized metricity condition $\tilde{\nabla}_{\mu}g^{\lambda\nu}=0$  if the vierbein obeys the generalized $\tilde{D}_{\mu}V^{a\lambda}=0$.

\subsection{Connections and the Dirac Equation}

To introduce spinors one starts with the free massless Dirac action in flat space, viz. the Poincare invariant $(1/2)\int d^4xi\bar{\psi}\gamma^{a}\partial_a\psi$ plus its Hermitian conjugate (or equivalently $(1/2)\int d^4xi\bar{\psi}\gamma^{a}\partial_a\psi $ plus its $CPT$ conjugate), where the fixed basis Dirac gamma matrices obey $\gamma_a\gamma_b+\gamma_b\gamma_a=2\eta_{ab}$ (with ${\rm diag}[\eta_{ab}]=(1,-1,-1,-1)$ here). To make this action invariant under local translations one introduces a $(-g)^{1/2}$ factor in the measure and replaces $\gamma^a\partial_a$ by $\gamma^aV^{\mu}_a\partial_{\mu}$, and to make the action locally Lorentz invariant one introduces the spin connection. Thus, in a standard curved Riemannian space with connections $\Lambda^{\lambda}_{\phantom{\alpha}\mu\nu}$ and $\omega_{\mu}^{ab}$, the Dirac action is given by 
\begin{eqnarray}
I_{\rm D}=\frac{1}{2}\int d^4x(-g)^{1/2}i\bar{\psi}\gamma^{a}V^{\mu}_a(\partial_{\mu}+\Sigma_{bc}\omega^{bc}_{\mu})\psi +H. c.,~
\label{16}
\end{eqnarray}
where $\Sigma_{ab}=(1/8)(\gamma_a\gamma_b-\gamma_b\gamma_a)$. Following an integration by parts and some algebraic steps $I_{\rm D}$ can be written as
\begin{eqnarray}
I_{\rm D}=\int d^4x(-g)^{1/2}i\bar{\psi}\gamma^{a}V^{\mu}_a(\partial_{\mu}+\Sigma_{bc}\omega^{bc}_{\mu})\psi.
\label{17}
\end{eqnarray}

As is familiar from experience with flat space actions, we see that the inclusion of the  Hermitian conjugate did not generate any new terms in the action. However, for connections more general than the Levi-Civita-based one, this is no longer the case. When one has a more general connection the Dirac action is given by 
\begin{eqnarray}
\tilde{I}_{\rm D}=\frac{1}{2}\int d^4x(-g)^{1/2}i\bar{\psi}\gamma^{a}V^{\mu}_a(\partial_{\mu}+\Sigma_{bc}\tilde{\omega}^{bc}_{\mu})\psi+H. c.~~
\label{18}
\end{eqnarray}
Following a few algebraic steps $\tilde{I}_{\rm D}$ is found to take the form 
\begin{eqnarray}
\tilde{I}_{\rm D}&=&I_{\rm D}+\frac{1}{16}\int d^4x(-g)^{1/2}i\bar{\psi}V^{a\mu}V^{b\lambda}V^{c \nu}
\nonumber\\
&\times&(\delta{\Gamma}_{\lambda\nu\mu}\gamma_a[\gamma_b,\gamma_c]+
(\delta{\Gamma}_{\lambda\nu\mu})^{\dagger}[\gamma_b,\gamma_c]\gamma_a)\psi
\label{19}
\end{eqnarray}
with some additional terms now being generated. It is these explicit additional  terms that will enable us to metricate the fundamental forces.  

With a view to what is to follow below, in (\ref{19})  we have expressly not taken $\delta{\Gamma}^{\lambda}_{\phantom{\alpha}\nu\mu}$ to be real or Hermitian. Recalling that 
\begin{eqnarray}
\gamma^a[\gamma^b,\gamma^c]-[\gamma^b,\gamma^c]\gamma^{a}&=&4\eta^{ab}\gamma^c-4\eta^{ac}\gamma^b
\nonumber\\
\gamma^a[\gamma^b,\gamma^c]+[\gamma^b,\gamma^c]\gamma^a&=&4i\epsilon^{abcd}\gamma_{d}\gamma^{5}, 
\nonumber\\
 \gamma^5&=&i\gamma^0\gamma^1\gamma^2\gamma^3,
\nonumber\\
\epsilon^{abcd}V^{\mu}_aV^{\nu}_bV^{\sigma}_cV^{\tau}_d&=&(-g)^{-1/2}\epsilon^{\mu\nu\sigma\tau},
\label{20}
\end{eqnarray}
we can rewrite $\tilde{I}_{\rm D}$ as
\begin{eqnarray}
\tilde{I}_{\rm D}=I_{\rm D}+\frac{1}{4}\int d^4x(-g)^{1/2}i\bar{\psi}\gamma_d\delta \Gamma^{d}\psi
\label{21}
\end{eqnarray}
where 
\begin{eqnarray}
\delta \Gamma^{d}&=&\frac{1}{2}[\delta{\Gamma}_{\mu\lambda\nu}+(\delta{\Gamma}_{\mu\lambda\nu})^{\dagger}](-g)^{-1/2}\epsilon^{\mu\lambda\nu\tau}i\gamma^5V^d_{\tau}
\nonumber\\
&+&\frac{1}{2}[\delta{\Gamma}_{\mu\lambda\nu}-(\delta{\Gamma}_{\mu\lambda\nu})^{\dagger}][g^{\mu\lambda}V^{d\nu}-g^{\mu\nu}V^{d\lambda}].
\label{22}
\end{eqnarray}

As we see, if $\delta{\Gamma}^{\lambda}_{\phantom{\alpha}\nu\mu}$ is in fact real, the only connection that could couple in $\tilde{I}_{\rm D}$ would be that part of it that is antisymmetric on all three of its indices. Thus of the two connections of interest to us only the torsion-dependent $K^{\lambda}_{\phantom{\alpha}\nu\mu}$ as evaluated with a real $Q^{\lambda}_{\phantom{\alpha}\nu\mu}$ could possibly couple to the fermion, with $W^{\lambda}_{\phantom{\alpha}\nu\mu}$ as evaluated with a real $A_{\mu}$ not being able to couple to the fermion at all, a result first noted in  \cite{Hayashi1977}.  Thus the Weyl connection as introduced by Weyl (viz. one with a real $A_{\mu}$) could not serve to metricate electromagnetism, and such an $A_{\mu}$ could not serve as the electromagnetic vector potential. As we will show below, we will rectify this by taking the Weyl connection not to be Hermitian at all but to be $PT$ symmetric instead, in consequence of which $A_{\mu}$ will be replaced by $iA_{\mu}$ in it.

For the torsion contribution to $\tilde{I}_{\rm D}$ with a real $Q^{\lambda}_{\phantom{\alpha}\nu\mu}$ evaluation is straightforward and yields (see e.g. \cite{Kibble1963},\cite{Shapiro2002})
\begin{eqnarray}
\tilde{I}_{\rm D}=\int d^4x(-g)^{1/2}i\bar{\psi}\gamma^{a}V^{\mu}_a(\partial_{\mu}+\Sigma_{bc}\omega^{bc}_{\mu}-i\gamma^5S_{\mu})\psi,~~
\label{23}
\end{eqnarray}
where 
\begin{eqnarray}
S^{\mu}&=&\frac{1}{8}(-g)^{-1/2}\epsilon^{\mu\alpha\beta\gamma}Q_{\alpha\beta\gamma},
\nonumber\\
-4(-g)^{-1/2}\epsilon_{\mu\alpha\beta\gamma}S^{\mu}&=&
Q_{\alpha\beta\gamma}+Q_{\gamma\alpha\beta}+Q_{\beta\gamma\alpha}.~~~
\label{24}
\end{eqnarray}
In the action $\tilde{I}_{\rm D}$  we note that even though the torsion is only antisymmetric on two of its indices, just as required the only components of the torsion that appear in its torsion-dependent $S^{\mu}$ term are the four that constitute that part of the torsion that is antisymmetric on all three of its indices. These four torsion components couple to the fermion via an axial vector current, and thus couple not to the electric current but to a magnetic current instead. A possible role for $S_{\mu}$ in electromagnetism as an axial vector potential was discussed in \cite{Mannheim2014b}, and we will return to the issue below. However before we do this, we need to discuss the relation between $PT$ symmetry, conformal symmetry, and Lorentz symmetry.

\section{$PT$, Lorentz, Conformal, and $CPT$ Symmetries}

\subsection{$PT$ Symmetry}

A $PT$ transformation differs from either a conformal transformation or a Lorentz transformation in two significant ways. First it is not a continuous transformation but a discrete one, and second it is not a linear transformation but through time reversal is an antilinear one. Its utility for physics was developed by Bender and collaborators \cite{Bender2007} following the discovery \cite{Bender1998} that the eigenvalues of the non-Hermitian Hamiltonian $H=p^2+ix^3$  were all real. As we thus see, while Hermiticity is sufficient to yield real eigenvalues it is not necessary. With the Hamiltonian $H=p^2+ix^3$ being $PT$ symmetric ($PxP^{-1}=-x$, $TiT^{-1}=-i$), and with $E^*$ being an eigenvalue of any $PT$-symmetric Hamiltonian $H$ if $E$ is an eigenvalue ($HPT|\psi\rangle=PTH|\psi\rangle=PTE|\psi\rangle=E^*PT|\psi\rangle$), it was recognized that one could also get real eigenvalues via $PT$ symmetry. Subsequently it was recognized that the key issue was not the reality of the eigenvalues themselves but of the secular equation $f(\lambda)=|H-\lambda I|$ that determines them, with it being shown  first that if $H$ is $PT$ symmetric then $f(\lambda)$ is a real function of $\lambda$ \cite{Bender2002}, and second that if $f(\lambda)$ is a real function of $\lambda$, then $H$ must possess a $PT$ symmetry  \cite{Bender2010}. Since a complex $f(\lambda)$ would require that at least one eigenvalue be complex, $PT$ symmetry was thus identified as being the necessary condition for reality of eigenvalues.

A benefit of $PT$ symmetry  is that with it one can make statements about the eigenvalues of a Hamiltonian just by checking its symmetry structure, not only without any need to determine whether or not the Hamiltonian is Hermitian (which requires studying its behavior at asymptotic spatial infinity to check whether one can drop surface terms in integrations by parts), but without even needing to solve for the eigenvalues at all. Moreover, with $PT$ being a symmetry, one can study the symmetry of every path in a path integral quantization, and thus without actually doing the integration one can know ahead of time that the Hamiltonian of the quantum theory that will result will be $PT$ symmetric if every path integral path is. Since path integral quantization is a completely c-number approach to quantization, it makes no reference to any Hilbert space at all and thus makes no reference to any quantum Hamiltonian at all. Rather, the path integral generates the Green's functions of the quantum theory, i.e. it generates matrix elements of quantum operators. It is only after constructing the Hilbert space in which those operators act could one then determine whether or  not the quantum Hamiltonian might be Hermitian. With $PT$ symmetry on the other hand one knows a lot about the quantum theory before even starting to evaluate the path integral. In the same way as working not with the Hamiltonian but with the action integral of the Lagrangian has always been beneficial for establishing the symmetry structure of a quantum theory, it is equally the case for $PT$ symmetry.

When a Hamiltonian is not Hermitian it is not appropriate to use the Dirac norm, since if $|R(t)\rangle$ is a right eigenstate of $H$ then $\langle R(t)|R(t)\rangle=\langle R(0)|e^{iH^{\dagger}t}e^{-iHt}|R(0)\rangle$ is not equal to $\langle R(0)|R(0)\rangle$, with the norm not being time independent. However, if instead of being Hermitian the Hamiltonian is $PT$ symmetric, then one should use a norm involving not the Dirac conjugate of $|R(t)\rangle$ but its $PT$ conjugate instead \cite{Bender2007}. If we introduce a left eigenstate $\langle L(t)|$ of $H$, then the appropriate $PT$ theory norm can be written  \cite{Mannheim2013a} as the time independent $\langle L(t)|R(t)\rangle=\langle L(0)|e^{iHt}e^{-iHt}|R(0)\rangle=\langle L(0)|R(0)\rangle$. In this way one can obtain unitary time evolution in theories with non-Hermitian Hamiltonians, with it being shown in \cite{Mannheim2013a} that $PT$ symmetry of a Hamiltonian is a both necessary and sufficient condition for unitary time evolution, with Hermiticity only being sufficient one. 

A further benefit of the $PT$ theory norm is that in cases where the Dirac norm $\langle R(t)|R(t)\rangle$ is found to be of negative ghost state form, a cause for this can be that the Hamiltonian is not Hermitian, with one then not being permitted to use the Dirac norm. Thus rather than signaling that a theory is not unitary, the presence of a negative Dirac norm could be signaling that one is not in a Hermitian theory and that one should not be using the Dirac norm at all, and in such a situation the propagator would be given not  by $\langle \Omega_R|T(\phi(x)\phi(x^{\prime}))|\Omega_R\rangle$ but by $\langle \Omega_L|T(\phi(x)\phi(x^{\prime}))|\Omega_R\rangle$ instead. There are two cases with negative Dirac norms that have been identified in the literature as being $PT$ theories, with both of their $\langle L(t)|R(t)\rangle$ norms then being found to be positive definite. The $PT$ norm has been found to be relevant  \cite{Bender2005} to the Lee model, and \cite{Bender2008a,Bender2008b}, \cite{Mannheim2011a,Mannheim2012a} to the conformal gravity theory that we shall encounter below.

\subsection{$PT$ Symmetry and the Lorentz Group}

While $PT$ symmetry is thus seen to be more general than Hermiticity, as stressed in \cite{Bender2007} it is also a physical requirement on a theory rather than the mathematical requirement that $H=H^{\dagger}$.  Indeed, both parity and time reversal symmetries are physical ones  that many theories possess, and in relativistic field theory properties of $PT$ invariance carry over to $CPT$ invariance in those cases where $PT$ is not a symmetry but $CPT$ is. As regards Poincare invariance, we note that the Hamiltonian is the generator of time translations regardless of whether or not it might be Hermitian. And as regards Lorentz invariance, we note that the Lorentz group has a PT extension. Specifically, under the combined $PT$ transformation $x_{\mu}$ transforms as $x_{\mu}\rightarrow -x_{\mu}$, with $PT$ thus being compatible with Lorentz invariance as $PT$ (but not $P$ or $T$ separately) treats all four components of $x_{\mu}$ equivalently \cite{footnote1}.

Moreover, there is an intimate connection between $PT$ symmetry and the structure of the irreducible representations of the Lorentz group. Consider for instance the standard $\mathbf{E}$ and $\mathbf{B}$  fields of electromagnetism. The $\mathbf{E}$  field is $P$ odd and $T$ even, to thus be $PT$ odd, while  the $\mathbf{B}$  field is $P$ even and $T$ odd, to thus be $PT$ odd also. Lorentz transformations that mix the $\mathbf{E}$  and $\mathbf{B}$  fields thus mix fields with the same  $PT$. Now the $\mathbf{E}$  and $\mathbf{B}$  fields transform according to the $D(1,0 )\oplus D(0,1)$ representation of the Lorentz group. However, this representation is reducible, with the irreducible components being the left- and right-handed $\mathbf{E} -i\mathbf{B}$  and $\mathbf{E} +i\mathbf{B}$. While irreducible under the Lorentz group, as we see under a $PT$ transformation $\mathbf{E} -i\mathbf{B} \rightarrow -(\mathbf{E} +i\mathbf{B})$ \cite{footnote2}. The six fields $\mathbf{E}$  and $\mathbf{B}$  while reducible under $SO(3,1)$ alone are thus irreducible under $SO(3,1)\times PT$. Exactly the same is true of the left- and right-handed fermions, which respectively transform as $D(1/2,0)$ and $D(0,1/2)$. They are reducible under $SO(3,1)$ but irreducible under $SO(3,1)\times PT$ \cite{footnote3}.

An analogous pattern occurs for the vector and axial vector currents. For the vector current $J^{\mu}=\bar{\psi}\gamma^{\mu}\psi$ we note that $J^0$ is $P$ even and $T$ even, to thus be $PT$ even, while $J^i$ is $P$ odd and $T$ odd, to thus be $PT$ even also. Since the vector current couples to $A_{\mu}$, $A_{\mu}$ is $PT$ even. For the axial vector current $K^{\mu}=\bar{\psi}\gamma^{\mu}\gamma^5\psi$ we note that $K^0$ is $P$ odd and $T$ even, to thus be $PT$ odd, while $K^i$ is $P$ even and $T$ odd, to thus be $PT$ odd also. Since $S_{\mu}$ couples to the axial current in the generalized Dirac action $\tilde{I}_{\rm D}$ given in (\ref{23}), it follows that $S_{\mu}$ is $PT$ odd.

\subsection{Global Conformal Symmetry}

As well as being able to relate left- and right-handed irreducible representations of the Lorentz group via a discrete $PT$ symmetry, it is also possible to relate them via a set of continuous transformations instead, with the requisite transformations being conformal transformations, viz. precisely those transformations that are relevant to the Weyl geometry of interest to us in this paper. In flat space the conformal group enlarges the 10 parameter flat space Poincare group with its  $P^{\mu}$ and $M^{\mu\nu}$ generators to include five more flat space generators, a dilatation operator $D$ and four conformal generators $C^{\mu}$.  With respective constant parameters $\epsilon^{\mu}$, $\Lambda^{\mu}_{\phantom{\mu}\nu}$, $\lambda$ and $c^{\mu}$ the 15 generators transform $x^{\mu}$ and $x^2$ according to 
\begin{eqnarray}
x^{\mu}&\rightarrow& x^{\mu}+\epsilon^{\mu},\qquad x^{\mu} \rightarrow \Lambda^{\mu}_{\phantom{\mu}\nu}x^{\nu}
\nonumber\\
x^{\mu}&\rightarrow& \lambda x^{\mu},\qquad x^{\mu} \rightarrow \frac{x^{\mu}+c^{\mu}x^2}{1+2c\cdot x+c^2x^2},
\nonumber \\
x^2&\rightarrow& \lambda^2x^2,\qquad x^2 \rightarrow \frac{x^2}{1+2c\cdot x+x^2}.
\label{25}
\end{eqnarray}
With the 15 infinitesimal generators acting on the coordinates $x^{\mu}$ according to ($\partial_{\mu}=(\partial/\partial t, \partial/\partial \bf{x})$, $\partial^{\mu}=(\partial/\partial t, -\partial/\partial \bf{x})$ here)
\begin{eqnarray}
P^{\mu}&=&i\partial^{\mu},\qquad M^{\mu\nu}=i(x^{\mu}\partial^{\nu}-x^{\nu}\partial^{\mu}),
\nonumber\\
D&=&ix^{\mu}\partial_{\mu}~~~~C^{\mu}=i(x^2\eta^{\mu\nu}-2x^{\mu}x^{\nu})\partial_{\nu},
\label{26}
\end{eqnarray}
together they form the 15-parameter $SO(4,2)$ conformal group, with algebra
\begin{eqnarray}
&&[M_{\mu\nu},M_{\rho\sigma}]=i(-\eta_{\mu\rho}M_{\nu\sigma}+\eta_{\nu\rho}M_{\mu\sigma}
\nonumber\\
&&~~~~~~~~~~~~~~~-\eta_{\mu\sigma}M_{\rho\nu}+\eta_{\nu\sigma}M_{\rho\mu}),
\nonumber\\
&&[M_{\mu\nu},P_{\sigma}]=i(\eta_{\nu\sigma}P_{\mu}-\eta_{\mu\sigma}P_{\nu}),~~~[P_{\mu},P_{\nu}]=0,~~~
\nonumber\\
&&[M_{\mu\nu},C_{\sigma}]=i(\eta_{\nu\sigma}C_{\mu}-\eta_{\mu\sigma}C_{\nu}),~~~
[M_{\mu\nu},D]=0,
\nonumber\\
&&[C_{\mu},C_{\nu}]=0,~~~[C_{\mu},P_{\nu}]=2i(\eta_{\mu\nu}D-M_{\mu\nu}),
\nonumber\\
&&[D,P_{\mu}]=-iP_{\mu},~~~[D,C_{\mu}]=iC_{\mu}.
\label{27}
\end{eqnarray}

The utility of the conformal group is that while timelike, lightlike or spacelike distances are preserved by the 10 Poincare transformations, lightlike distances are preserved by all 15 conformal group transformations, with the  light cone thus having  a symmetry larger than Poincare. With the flat space free massless particle propagator also depending only on the distance (cf. $1/x^2$ for spin zero scalars and $\gamma_{\mu}x^{\mu}/x^4$ for spin one half fermions), free flat space massless particles possess all 15 conformal group invariances. Theories in which all particles are massless at the level of the Lagrangian and all coupling constants are dimensionless thus have an underlying conformal structure. With conformal invariance being tied in with masslessness at the level of the Lagrangian, to generate masses we would thus have to break the conformal symmetry via vacuum dynamics. Moreover, this is precisely the standard $SU(3)\times SU(2)\times U(1)$ picture of strong, electromagnetic and weak interactions, where all fermions and gauge bosons have no mass at the level of the Lagrangian and all couplings in the pure fermion gauge boson sector  are dimensionless. When we make the conformal transformations local, which we do below, this will lead us to a theory of gravity, conformal gravity (a strictly Riemnannian variant of the Weyl geometry of interest to us in this paper), in which its coupling constants are dimensionless too. 

The conformal algebra admits of a 4-dimensional spinor representation since the 15 Dirac matrices $\gamma^5$, $\gamma^{\mu}$, $\gamma^{\mu}\gamma^5$, $[\gamma^{\mu},\gamma^{\nu}]$ also close on the $SO(4,2)$ algebra. The group $SU(2,2)$ is the covering group of $SO(4,2)$ with the 4-dimensional spinor being its fundamental representation. Thus unlike the Lorentz group $SO(3,1)$ where a 4-component spinor transforms according to the $D(1/2,0) \oplus D(0,1/2)$ representation, under the conformal group all four components are irreducible, with the conformal transformations mixing the left- and right-handed spinors, doing so via  transformations that are continuous. Since this holds for all spinors no matter what their internal quantum numbers might be, in a conformal invariant theory  neutrinos would have to have four components too, with right-handed neutrinos being needed to accompany the observed left-handed ones.

The fact that 4-component fermions are irreducible under the conformal group means that conformal transformations mix components with opposite $PT$. In \cite{footnote3} we had noted that under a $PT$ transformation a Dirac spinor  transforms as $PT\psi(t,\mathbf{x})T^{-1}P^{-1}=-\gamma^2\gamma^5\psi(-t,-\mathbf{x})$, with its conjugate transforming as $PT\bar{\psi}(t,\mathbf{x})T^{-1}P^{-1}=-\bar{\psi}(-t,-\mathbf{x})\gamma^2\gamma^5$. We now recognize this transformation as being none other than a conformal transformation since $\gamma^2\gamma^5$ is one of the 15 generators of the conformal group.  $PT$ symmetry is thus integrally connected with conformal symmetry. And because of this, conformal transformations will thus mix Lorentz group representations such as $\mathbf{E} -i\mathbf{B}$ and $\mathbf{E} +i\mathbf{B}$. 

Given the fundamental 4-dimensional representation of the conformal group, by constructing the $4\otimes 4^*$ direct product we can make both a 15-dimensional adjoint representation of the conformal group and a singlet. With the 15 Dirac gamma matrices and the identity matrix spanning a general $4\times 4$ matrix space, we see that in the irreducible decomposition of $4\otimes 4^*$ we have precisely the needed number of independent Dirac gamma matrices. We can thus anticipate that the associated fermion bilinear currents $\bar{\psi}\Gamma\psi$ with $\Gamma=1,i\gamma^5,\gamma^{\mu},\gamma^{\mu}\gamma^5,i[\gamma^{\mu},\gamma^{\nu}]$ could play a central role in physics, with  $\Gamma=\gamma^{\mu}$ and $\Gamma=\gamma^{\mu}\gamma^5$ being seen to appear in $\tilde{I}_{\rm D}$ (cf. (\ref{21})) or $\tilde{J}_{\rm D}$, its $A_{\mu}$ extension given in (\ref{33}) below \cite{footnote4}.

\subsection{Local Conformal Symmetry}

In order to extend the above global conformal symmetry to a local  symmetry, we note that while the conformal group has 15 generators no 4-dimensional space can have more than 10 Killing vectors, viz. vectors that obey $\nabla_{\mu}K_{\nu}+\nabla_{\nu}K_{\mu}=0$. Since flat spacetime is maximally 4-symmetric it has 10 vectors that obey  $\partial_{\mu}K_{\nu}+\partial_{\nu}K_{\mu}=0$, viz. the 10 $K_{\mu}$ that are embodied in
\begin{eqnarray}
K_{\mu}=a_{\mu}+b_{\mu\nu}x^{\nu},
\label{28}
\end{eqnarray}
where $a_{\mu}$ is a constant four-vector and $b_{\mu\nu}$ is a constant 6-component antisymmetric rank two tensor. To account for the remaining five generators of the conformal group we introduce conformal Killing vectors, viz. vectors that obey $\nabla_{\mu}K_{\nu}+\nabla_{\nu}K_{\mu}=f(x)g_{\mu\nu}$ where $f(x)$ is an appropriate scalar function. For flat spacetime  we find that with $\lambda$ being a constant scalar and $c_{\mu}$ being a constant four-vector the five $K_{\mu}$ that are embodied in 
\begin{eqnarray}
K_{\mu}=\lambda x_{\mu}+c_{\mu}x^2-2x_{\mu}c\cdot x
\label{29}
\end{eqnarray}
obey
\begin{eqnarray}
\partial_{\mu}K_{\nu}+\partial_{\nu}K_{\mu}=2(\lambda-2c\cdot x)\eta_{\mu\nu}.
\label{30}
\end{eqnarray}
If we now allow $a_{\mu}$, $b_{\mu\nu}$, $\lambda$, and $c_{\mu}$ to become spacetime dependent, we note that  the $\lambda-2c\cdot x$ factor in (\ref{30}) becomes just one general spacetime-dependent function. We thus anticipate having only 11 local symmetries rather than the initial 15 global ones. (This is to be expected since under the global $D$ and $C_{\mu}$ transformations given in (\ref{25}) $x^2$ transforms as $x^2\rightarrow \lambda^2x^2$, $x^2 \rightarrow x^2/(1+2c\cdot x+x^2)$, with a global $C_{\mu}$ being a particular local $D$. Also, as can be seen from (\ref{27}), $M_{\mu\nu}$, $P_{\mu}$ and $D$ close on an algebra all on their own.) Referring now to the Dirac action $I_{\rm D}=\int d^4x(-g)^{1/2}i\bar{\psi}\gamma^{a}V^{\mu}_a(\partial_{\mu}+\Sigma_{bc}\omega^{bc}_{\mu})\psi$ given in (\ref{17}), we see that it possesses four local translation invariances and six local Lorentz invariances (as it of course must since the $V^{\mu}_a$ vierbeins and the $\omega^{bc}_{\mu}$ spin connection were expressly introduced for this purpose). However, $I_{\rm D}$ also possesses one local conformal invariance as well, since it is left invariant under $g_{\mu\nu}(x)\rightarrow e^{2\alpha(x)}g_{\mu\nu}(x)$, $V^a_{\mu}(x)\rightarrow e^{\alpha(x)}V^a_{\mu}(x)$, $\psi(x)\rightarrow e^{-3\alpha(x)/2}\psi(x)$ with arbitrary spacetime dependent $\alpha(x)$. With this local conformal invariance we find that $I_{\rm D}$ does indeed have 11 local invariances \cite{footnote5}, just as anticipated \cite{footnote6}.

As had been noted above, we could generalize  $I_{\rm D}$ to the general $\tilde{I}_{\rm D}$ given in (\ref{19}) provided $\delta{\Gamma}^{\lambda}_{\phantom{\alpha}\nu\mu}$ is itself a true rank-three tensor. For any $\delta{\Gamma}^{\lambda}_{\phantom{\alpha}\nu\mu}$ that is a rank-three tensor the general $\tilde{I}_{\rm D}$ will still be both locally translation invariant and locally Lorentz invariant. However, requiring that the contribution of $\delta{\Gamma}^{\lambda}_{\phantom{\alpha}\nu\mu}$ to $\tilde{I}_{\rm D}$ also be locally conformal invariant will constrain how the fields in  $\delta{\Gamma}^{\lambda}_{\phantom{\alpha}\nu\mu}$ are to transform under a local conformal transformation.

To therefore identify the conformal properties needed for the $S_{\mu}$ term in $\tilde{I}_{\rm D}$, we note that since the Levi-Civita connection transforms as 
\begin{eqnarray}
\Lambda^{\lambda}_{\phantom{\sigma}\mu\nu}\rightarrow \Lambda^{\lambda}_{\phantom{\sigma}\mu\nu}
+(\delta^{\lambda}_{\mu}\partial_{\nu}+\delta^{\lambda}_{\nu}\partial_{\mu}-g_{\mu\nu}\partial^{\lambda})\alpha(x),~~
\label{31}
\end{eqnarray}
a straightforward transformation for the torsion that takes into account its antisymmetry structure is \cite{Buchbinder1985,Shapiro2002}
\begin{eqnarray}
Q^{\lambda}_{\phantom{\sigma}\mu\nu}\rightarrow Q^{\lambda}_{\phantom{\sigma}\mu\nu}
+q(\delta^{\lambda}_{\mu}\partial_{\nu}-\delta^{\lambda}_{\nu}\partial_{\mu})\alpha(x),
\label{32}
\end{eqnarray}
where $q$ is the conformal weight of the torsion tensor. While the specific value taken by  $q$ is not known, there appear to be two natural choices for it. One of course is simply $q=0$. And since the torsion tensor has to have the same engineering dimension as the Levi-Civita connection, it must have engineering dimension equal to one, with $q=1$ thus being the other. However regardless of this, it was noted in \cite{Fabbri2014} that in fact no matter what the value of $q$, the $\alpha(x)$-dependent term in (\ref{32}) actually drops out identically in $S_{\mu}$, with $S_{\mu}$ thus having conformal weight equal to zero.  Since the term that $S_{\mu}$ couples to in $\tilde{I}_{\rm D}$ , viz. $(-g)^{1/2}\bar{\psi}\gamma^{a}V^{\mu}_a\gamma^5\psi$, has conformal weight zero itself ($4-3/2-1-3/2=0$), we thus establish that the $S_{\mu}$-dependent term in $\tilde{I}_{\rm D}$ term is locally conformal invariant, just as required.

There is, however, a completely different local way to view the $S_{\mu}$-dependent term in  $\tilde{I}_{\rm D}$. Suppose we start with the torsion independent $I_{\rm D}$ and instead of changing the connection at all require that the action be invariant under  a local chiral transformation on the fermion of the form $\psi(x)\rightarrow e^{i\gamma^5\beta(x)}\psi(x)$ with spacetime-dependent $\beta(x)$. To maintain the chiral symmetry we would need to minimally couple in an axial vector field $S_{\mu}(x)$ that transforms as $S_{\mu}(x)\rightarrow S_{\mu}(x)+\partial_{\mu}\beta(x)$, and the resulting action that we would obtain would be precisely none other than $\tilde{I}_{\rm D}$ as given in (\ref{23}). In such a case we would have to appeal to the zero conformal weight of $(-g)^{1/2}\bar{\psi}\gamma^{a}V^{\mu}_a\gamma^5\psi$  to establish that $S_{\mu}(x)$ should have conformal weight zero itself \cite{footnote7}. While inspection of $\tilde{I}_{\rm D}$ alone could thus not tell us whether $S_{\mu}(x)$ is associated with a torsionless geometry or with one with torsion, the geometry would still know, since one would have to use either $R^{\lambda}_{\phantom{\rho}\mu\nu\kappa}$ or $\tilde{R}^{\lambda}_{\phantom{\rho}\mu\nu\kappa}$. However, as we will see below, even this distinction will disappear; and even if there were to be a distinction, it itself would not involve any parallel transport problem since the covariant derivative $\tilde{\nabla}_{\mu}g^{\lambda\nu}$ of the metric as constructed with the contorsion  tensor $K^{\lambda}_{\phantom{\alpha}\mu\nu}$ is zero.

In the same way that we could introduce $S_{\mu}$ via a local axial symmetry, we could equally of course introduce the vector potential $A_{\mu}$ via a local vector symmetry, since on requiring invariance under $\psi(x)\rightarrow e^{i\alpha(x)}\psi(x)$ with spacetime-dependent $\alpha(x)$ we would need to minimally couple in a vector field $A_{\mu}$ that transforms as $A_{\mu}(x)\rightarrow A_{\mu}(x)+\partial_{\mu}\alpha(x)$. With such a coupling $\tilde{I}_{\rm D}$ would be replaced by 
\begin{eqnarray}
\tilde{J}_{\rm D}&=&\int d^4x(-g)^{1/2}i\bar{\psi}\gamma^{a}V^{\mu}_a(\partial_{\mu}+\Sigma_{bc}\omega^{bc}_{\mu}
\nonumber\\
&&-iA_{\mu} -i\gamma^5S_{\mu})\psi.
\label{33}
\end{eqnarray}
Given that $(-g)^{1/2}\bar{\psi}\gamma^{a}V^{\mu}_a\psi$ has conformal weight zero, $\tilde{J}_{\rm D}$ will be locally conformal invariant if, just like the axial $S_{\mu}$, the vector $A_{\mu}$ has conformal weight zero too. In fact just as had been discussed in \cite{footnote7} in regard to $S_{\mu}$, the conformal weight of $A_{\mu}$ can also be determined from global scale invariance considerations alone. The fact that $A_{\mu}$ is not to transform under a local conformal transformation is of significant import since it constitutes a quite major departure from Weyl's original intent that it is to transform non-trivially under a conformal transformation, a point that will prove crucial below.

Regardless of how it may or may not have been derived, as an action $\tilde{J}_{\rm D}$ is quite remarkable as it has a very rich local invariance structure. $\tilde{J}_{\rm D}$ is invariant under local translations, local Lorentz transformations, local gauge transformations, local axial gauge transformations, and local conformal transformations. Moreover, $\tilde{J}_{\rm d}$ is not just invariant under any arbitrary set of local transformations, it is invariant under some of the key local transformations in physics \cite{footnote8}. 

\subsection{$PT$ Symmetry and $CPT$ Symmetry}

Beyond all these continuous symmetries, $\tilde{J}_{\rm D}$ has two further symmetries, as  it is invariant under a discrete $PT$ symmetry and a discrete $CTP$ symmetry. As regards first $PT$ symmetry, we note that given the $PT$ transformation properties of the fermion fields, the generic $\bar{\psi}(x^{\mu})\Gamma\psi(x^{\mu})$ will transform into $\bar{\psi}(-x^{\mu})\gamma^2\gamma^5\Gamma^*\gamma^2\gamma^5\psi(-x^{\mu})$. Thus  $\bar{\psi}\psi$,  $\bar{\psi}i\gamma^5\psi$, and $\bar{\psi}\gamma^{\mu}\psi$ are $PT$ even, while $\bar{\psi}\gamma^{\mu}\gamma^5\psi$ and $\bar{\psi}i[\gamma^{\mu},\gamma^{\nu}]\psi$ are $PT$ odd. Now we had noted earlier that $A_{\mu}$ is $PT$ even and $S_{\mu}$ is $PT$ odd. With $i\partial_{\mu}$ and $iw^{bc}_{\mu}$ both being $PT$ even [in (\ref{14}) the $[\partial/\partial x^{\mu}]V^{a\nu}(-x^{\mu})=-[\partial/\partial (-x^{\mu})]V^{a\nu}(-x^{\mu})$ term in  $V^b_{\nu}\partial_{\mu}V^{a\nu}$, and analogously for $\Lambda^{\lambda}_{\phantom{\lambda}\nu\mu}$, makes $w^{bc}_{\mu}$ act as an odd $PT$ operator in the $\int d^4x=\int d^4(-x)$ integration], we see that every term in $\tilde{J}_{\rm D}$ is $PT$ even. $PT$ symmetry is thus again seen to accompany conformal symmetry.  

As regards $CPT$ symmetry, we recall that, with $C\psi(t,\mathbf{x})C^{-1}=i\gamma^2\psi^{\dagger}(t,\mathbf{x})$, $C\bar{\psi}(t,\mathbf{x})C^{-1}=\psi(t,\mathbf{x})i\gamma^2\gamma^0$, under $CPT$ we obtain $CPT\psi(t,\mathbf{x})T^{-1}P^{-1}C^{-1}=i\psi^{\dagger}(-t,-\mathbf{x})\gamma^5$, $CPT\bar{\psi}(t,\mathbf{x})T^{-1}P^{-1}C^{-1}=-i\gamma^0\gamma^5\psi(-t,-\mathbf{x})$.  Following an antisymmetric interchange of $\psi$ and $\bar{\psi}$ both $\bar{\psi}\gamma^{\mu}\psi$ and  $\bar{\psi}\gamma^{\mu}\gamma^5\psi$ are found to be $CPT$ odd, while $\bar{\psi}\psi$, $\bar{\psi}i\gamma^5 \psi$, and $\bar{\psi}i[\gamma^{\mu},\gamma^{\nu}]\psi$ are CPT even. Under the same antisymmetric interchange  we obtain
$(1/2)\int d^4x(-g)^{1/2}i\bar{\psi}\gamma^{a}V^{\mu}_a(\partial_{\mu}+\Sigma_{bc}\omega^{bc}_{\mu})\psi=-(1/2)\int d^4x(-g)^{1/2}i(\partial_{\mu}\psi)V^{\mu}_a(\gamma^{a})^{\rm Tr}\bar{\psi}-(1/2)\int d^4x(-g)^{1/2}i\psi V^{\mu}_a(\Sigma_{bc})^{\rm Tr}\omega^{bc}_{\mu}(\gamma^{a})^{\rm Tr}\bar{\psi}$ ($\rm{Tr}$ denotes transpose), to find that   the $CPT$ and Hermitian conjugates of this expression are equal. We thus establish that $I_{\rm D}$ as given in (\ref{16}) is $CPT$ symmetric. With both $\bar{\psi}V^{\mu}_a\gamma^a\psi$ and $\bar{\psi}V^{\mu}_a\gamma^a\gamma^5\psi$  being $CPT$ odd, the full $\tilde{J}_{\rm D}$ is $CPT$ invariant since $A^{\mu}$ and $S^{\mu}$ are both $CPT$ odd also ($A_{\mu}$ is  $PT$ even and $C$ odd, and $S_{\mu}$ is $PT$ odd and $C$ even). Minimal coupling is thus fullly $CPT$ symmetric.

For  the contribution of $\delta{\omega}^{bc}_{\mu}=\tilde{\omega}^{bc}_{\mu}-\omega^{bc}_{\mu}$, we note that the Hermitian and $CPT$ conjugates of $(1/2)\int d^4x(-g)^{1/2}i\bar{\psi}\gamma^{a}V^{\mu}_a\Sigma_{bc}\delta{\omega}^{bc}_{\mu})\psi$ are given by
$(1/2)\int d^4x(-g)^{1/2}i\bar{\psi}V^{\mu}_a(\delta{\omega}^{bc}_{\mu})^{\dagger}\Sigma_{bc}\gamma^{a}\psi$ and $-(1/2)\int d^4x(-g)^{1/2}i\bar{\psi}V^{\mu}_a(\delta{\omega}^{bc}_{\mu})^{CPT}\Sigma_{bc}\gamma^{a}\psi$. These two conjugates will thus coincide if $(\delta{\omega}^{bc}_{\mu})^{\dagger}=-(\delta{\omega}^{bc}_{\mu})^{CPT}$, but not otherwise. With a metricated $S_{\mu}$ (and thus $Q_{\alpha\beta\gamma}$) being Hermitian and $CPT$ odd, and with the $iA_{\mu}$-based connection that we actually use below being  anti-Hermitian and $CPT$ even, the metrication of both $A_{\mu}$ and $S_{\mu}$ studied in this paper is thus fully compatible with both $PT$ symmetry and $CPT$ symmetry.

Given all of these remarks, we see that in general if we wish to consider any specific contribution to the generic connection $\delta{\Gamma}^{\lambda}_{\phantom{\alpha}\nu\mu}$, each such  contribution is constrained in three distinct ways. The contribution to $\delta{\Gamma}^{\lambda}_{\phantom{\alpha}\nu\mu}$ would need to be a true rank-three tensor, it would need to keep $\tilde{I}_{\rm D}$ locally conformal invariant, and it would need to keep $\tilde{I}_{\rm D}$ $PT$ (and also $CPT$) even. Since we have seen that we can introduce $S_{\mu}$ either by a local gauge invariance or by a metrication that meets these three requirements, it is natural to ask whether we could do the same for $A_{\mu}$ and introduce it by a metrication procedure that meets these three requirements as well. However in order to do so for the electromagnetic vector potential $A_{\mu}$, we first need to discuss the relation of the axial $S_{\mu}$ to electromagnetism.

\section{The Relation of $S_{\mu}$ to Electromagnetism}

\subsection{An Axial Vector Potential}

While we have related $S_{\mu}$ to torsion in the above, $S_{\mu}$ can also be related to electromagnetism.  If we consider the standard Maxwell equations as coupled to an electric vector current $J^{\mu}$ in a standard curved Riemannian background geometry, viz.
\begin{eqnarray}
\nabla_{\nu}F^{\nu\mu}=J^{\mu},\qquad (-g)^{-1/2}\epsilon^{\mu\nu\sigma\tau}\nabla_{\nu}F_{\sigma\tau}
=0,
\label{34}
\end{eqnarray}
we count a total of eight equations. If we wish to obtain all eight of these equations via a variational principle we would need to vary with respect to eight different quantities \cite{footnote9}. As noted in \cite{Mannheim2014b}, given the structure of (\ref{34}) these eight would need to be a vector $A_{\mu}$ and an axial vector $S_{\mu}$. In fact one should use these eight potentials if magnetic currents are present.  Indeed, recalling the study \cite{Shanmugadhasan1952,Cabibbo1962} of the magnetic monopole problem, it is very convenient to introduce 
\begin{eqnarray}
X^{\mu\nu}&=&\nabla^{\mu}A^{\nu}-\nabla^{\nu}A^{\mu}
\nonumber\\
&-&\frac{1}{2}(-g)^{-1/2}\epsilon^{\mu\nu\sigma\tau}(\nabla_{\sigma}S_{\tau}-\nabla_{\tau}S_{\sigma})
\label{35}
\end{eqnarray}
as a generalized $F^{\mu\nu}$. On setting $S^{\mu\nu}=\nabla^{\mu}S^{\nu}-\nabla^{\nu}S^{\mu}$, we can rewrite $X^{\mu\nu}$ in terms of the standard $F^{\mu\nu}=\nabla^{\mu}A^{\nu}-\nabla^{\nu}A^{\mu}$ and the dual $\hat{S}^{\mu\nu}=(1/2)(-g)^{-1/2}\epsilon^{\mu\nu\sigma\tau}S_{\sigma\tau}$ of $S^{\mu\nu}$ according to:
\begin{eqnarray}
X^{\mu\nu}=F^{\mu\nu}-\hat{S}^{\mu\nu},~~~\hat{X}^{\mu\nu}=\hat{F}^{\mu\nu}+S^{\mu\nu}.
\label{36}
\end{eqnarray}
(If $\epsilon^{0123}=+1$, $\epsilon_{0123}=-1$.)
Given this $X^{\mu\nu}$, (\ref{34}) is replaced by 
\begin{eqnarray}
\nabla_{\nu}X^{\nu\mu}&=&\nabla_{\nu}F^{\nu\mu}=J^{\mu},\qquad \nabla_{\nu}\hat{X}^{\nu\mu}=\nabla_{\nu}S^{\nu\mu}=K^{\mu},
\nonumber\\
\nabla_{\nu}\hat{F}^{\nu\mu}&=&0,\qquad \nabla_{\nu}\hat{S}^{\nu\mu}=0,
\label{37}
\end{eqnarray}
where $K^{\mu}$ is a magnetic current, with it being $\nabla_{\nu}\hat{X}^{\nu\mu}=K^{\mu}$ that is to describe the magnetic monopole sector.

On introducing the action 
\begin{eqnarray}
I=\int d^4x(-g)^{1/2}\bigg{[} -\frac{1}{4}X_{\mu\nu}X^{\mu\nu}-A_{\mu}J^{\mu}-S_{\mu}K^{\mu}\bigg{]},~~
\label{38}
\end{eqnarray}
we find that stationary variation with respect to $A_{\mu}$ and $S_{\mu}$ then immediately leads to  (\ref{37}), just as we would want. Moreover, up to surface terms this action decomposes into two sectors according to 
\begin{eqnarray}
I&=&\int d^4x(-g)^{1/2}\bigg{[} -\frac{1}{4}F_{\mu\nu}F^{\mu\nu}-A_{\mu}J^{\mu}
\nonumber\\
&-&\frac{1}{4}S_{\mu\nu}S^{\mu\nu}-S_{\mu}K^{\mu}\bigg{]}.
\label{39}
\end{eqnarray}
with the $A_{\mu}$ and $S_{\mu}$ sectors thus being decoupled in the action. Inspection of (\ref{39}) shows it to be both locally conformal invariant and PT symmetric, again just as we would want \cite{footnote10}.

With the usual $F^{01}=-E_x$, $F^{12}=-B_z$ etc. identification of the field strengths, we can give physical significance to the $S^{\mu}$ sector by introducing a second set of field strengths $S^{01}=-B_x^{\prime}$, $S^{12}=+E_z^{\prime}$, $\hat{S}^{01}=E_x^{\prime}$, $\hat{S}^{12}=B_z^{\prime}$, etc. In terms of the field strengths, we find that in flat space with $J^{\mu}=(\rho_e,\mathbf{J}_e)$ and $K^{\mu}=(\rho_m,-\mathbf{J}_m)$, the generalized Maxwell equations given in (\ref{37}) decompose into the standard sector
\begin{eqnarray} 
&&\boldsymbol{\nabla} \times \boldsymbol{B}-\frac{\partial\boldsymbol{E}}{\partial t}=\boldsymbol{J}_e,\qquad \boldsymbol{\nabla}\cdot\boldsymbol{E}=\rho_e,
\nonumber\\
&& \boldsymbol{\nabla}\times \boldsymbol{E}+\frac{\partial\boldsymbol{B}}{\partial t}=0,\qquad \boldsymbol{\nabla}\cdot\boldsymbol{B}=0,
\label{40}
\end{eqnarray}
and a primed sector 
\begin{eqnarray} 
&&\boldsymbol{\nabla} \times \boldsymbol{B}^{\prime}-\frac{\partial\boldsymbol{E}^{\prime}}{\partial t}=0,\qquad \boldsymbol{\nabla}\cdot\boldsymbol{E}^{\prime}=0,
\nonumber\\
&& \boldsymbol{\nabla}\times \boldsymbol{E}^{\prime}+\frac{\partial\boldsymbol{B}^{\prime}}{\partial t}=\boldsymbol{J}_m,\qquad \boldsymbol{\nabla}\cdot\boldsymbol{B}^{\prime}=\rho_m.
\label{41}
\end{eqnarray}
Finally, if we define $\mathbf{E}_{\rm TOT}=\mathbf{E}+\mathbf{E}^{\prime}$, $\mathbf{B}_{\rm TOT}=\mathbf{B}+\mathbf{B}^{\prime}$, we can combine  (\ref{40}) and (\ref{41}) into 
\begin{eqnarray} 
&&\boldsymbol{\nabla} \times \boldsymbol{B}_{\rm TOT}-\frac{\partial\boldsymbol{E}_{\rm TOT}}{\partial t}=\boldsymbol{J}_e,~~\boldsymbol{\nabla}\cdot\boldsymbol{E}_{\rm TOT}=\rho_e,~~~~~
\nonumber\\
&& \boldsymbol{\nabla}\times \boldsymbol{E}_{\rm TOT}+\frac{\partial\boldsymbol{B}_{\rm TOT}}{\partial t}=\boldsymbol{J}_m,~~\boldsymbol{\nabla}\cdot\boldsymbol{B}_{\rm TOT}=\rho_m.~~~~~
\label{42}
\end{eqnarray}
Thus even if $\mathbf{J}_m$ and $\rho_m$ can be neglected, it is $\mathbf{E}_{\rm TOT}$ and $\mathbf{B}_{\rm TOT}$ that are measured in electromagnetic experiments.

\subsection{$PT$ Structure of Chiral Electromagnetism}

In terms of $P$, $T$ assignments, $K^0=\rho_m$ is $P$ odd and $T$ even, to thus be $PT$ odd, while $K^i=-J^i_m$ is $P$ even and $T$ odd, to thus be $PT$ odd also. Consequently, the $\mathbf{E}^{\prime}$  field is $P$ odd and $T$ odd, to thus be $PT$ even, while  the $\mathbf{B}^{\prime}$  field is $P$ even and $T$ even, to thus be $PT$ even also. With $\mathbf{E}$  and $\mathbf{B}$ both being $PT$ odd, we see that $\mathbf{E}_{\rm TOT}$ and $\mathbf{B}_{\rm TOT}$ contain components with opposite $PT$. However, no transition between them could be generated by the action given in (\ref{39}) since in it the $A_{\mu}$ and $S_{\mu}$ sectors are decoupled. To obtain any such transitions we could  introduce  the conformal invariant, $CPT$ invariant  couplings $A_{\mu}K^{\mu}$ and $S_{\mu}J^{\mu}$, though $PT$ symmetry would then be lost. The higher order coupling $A_{\mu}K^{\mu}S_{\nu}J^{\nu}$ is both $PT$ and $CPT$ invariant.

We summarize the discrete  transformation properties of the fields and currents of interest to us in a table
\begin{eqnarray}
\begin{array}{c|ccccccc|cccc}
&P&T&PT&CPT&&&&P&T&PT&CPT\\ 
\hline
\boldsymbol{E}&-&+&-&+&&&\boldsymbol{E}^{\prime}&-&-&+&+\\ 
\boldsymbol{B}&+&-&-&+&&&\boldsymbol{B}^{\prime}&+&+&+&+\\
\rho_e&+&+&+&-&&&\rho_m&-&+&-&-\\
\boldsymbol{J}_e&-&-&+&-&&&\boldsymbol{J}_m&+&-&-&-\\
A_0&+&+&+&-&&&S_0&-&+&-&-\\
\boldsymbol{A}&-&-&+&-&&&\boldsymbol{S}&+&-&-&-\\
\boldsymbol{\nabla}\cdot\boldsymbol{B}&-&-&+&-&&&\boldsymbol{\nabla}\cdot\boldsymbol{B}^{\prime}&-&+&-&-\\
\boldsymbol{\nabla}\cdot\boldsymbol{E}&+&+&+&-&&&\boldsymbol{\nabla}\cdot\boldsymbol{E}^{\prime}&+&-&-&-
\end{array}
\nonumber
\end{eqnarray}
in which we have also listed the properties of $\boldsymbol{\nabla}\cdot\boldsymbol{B}$, $\boldsymbol{\nabla}\cdot\boldsymbol{B}^{\prime}$, $\boldsymbol{\nabla}\cdot\boldsymbol{E}$, and $\boldsymbol{\nabla}\cdot\boldsymbol{E}^{\prime}$. As we see, only $\boldsymbol{\nabla}\cdot\boldsymbol{B}^{\prime}$ could couple to $\rho_m$, and only $\boldsymbol{\nabla}\cdot\boldsymbol{E}$ could couple to $\rho_e$. The primed sector  $\mathbf{B}^{\prime}$ is thus needed to provide a coupling to a magnetic monopole $\rho_m$ that $\bf{B}$ itself could not provide.

As introduced above $S_{\mu}$ is just an axial vector potential to be used in Maxwell theory, and does not  need to possess any relation to the $S_{\mu}$ that appears in the fermionic $\tilde{J}_{\rm D}$ given in (\ref{33}). To establish a relation we recall that when one does a $\tilde{J}_{\rm D}$ path integration  $\int D\bar{\psi}D\psi\exp(i\tilde{J}_{\rm D})$ over the fermions (equivalent to a one fermion loop Feynman graph) one generates   \cite{tHooft2010a},  \cite{Shapiro2002} an effective action of the form 
\begin{eqnarray}
I_{\rm EFF}&=&\int d^4x(-g)^{1/2}C\bigg{[}\frac{1}{20}\left[R_{\mu\nu}R^{\mu\nu}-\frac{1}{3}(R^{\alpha}_{\phantom{\alpha}\alpha})^2\right]
\nonumber\\
&+&\frac{1}{3}(\partial_{\mu}A_{\nu}-\partial_{\nu}A_{\mu})(\partial^{\mu}A^{\nu}-\partial^{\nu}A^{\mu})
\nonumber\\
&+&\frac{1}{3}(\partial_{\mu}S_{\nu}-\partial_{\nu}S_{\mu})(\partial^{\mu}S^{\nu}-\partial^{\nu}S^{\mu})\bigg{]},
\nonumber\\
&=&\int d^4x(-g)^{1/2}C\bigg{[}\frac{1}{20}\left[R_{\mu\nu}R^{\mu\nu}-\frac{1}{3}(R^{\alpha}_{\phantom{\alpha}\alpha})^2\right]
\nonumber\\
&+&\frac{1}{3}X_{\mu\nu}X^{\mu\nu}\bigg{]}
\label{43}
\end{eqnarray}
where $C$ is a  log divergent constant, $R_{\mu\nu}$ is the standard  Levi-Civita-based, torsionless, Ricci tensor, and $X_{\mu\nu}$ is as given in (\ref{35}). The action $I_{\rm EFF}$ possesses all the  local symmetries possessed by $\tilde{J}_{\rm D}$, with the appearance of the strictly Riemannian $R_{\mu\nu}R^{\mu\nu}-(1/3)(R^{\alpha}_{\phantom{\alpha}\alpha})^2$ term being characteristic of a gravity theory that is locally conformal invariant (see e.g. \cite{Mannheim2006,Mannheim2011a,Mannheim2012a}). Comparing now with (\ref{39}), we see that, up to renormalization constants, the action $I_{\rm EFF}$ is precisely of the form needed for Maxwell theory, with torsion thus providing a natural origin for the second potential that Maxwell theory needs \cite{footnote11}. 

\subsection{The Key Role of the Fermion}

In our work the fermionic action plays a central role. If we start with the free massless Dirac action in flat space, viz. the Poincare invariant $(1/2)\int d^4xi\bar{\psi}\gamma^{a}\partial_a\psi +H. c.$, then it is natural to introduce $A_{\mu}$ via a local vector gauge invariance, with standard QED being set up this way. However, starting from the same action it is just as natural to equally introduce $S_{\mu}$ via a local axial gauge invariance, with a chiral QED then being set up. That this option is not ordinarily followed is because QED is ordinarily discussed without consideration either of setting up a variational procedure for Faraday's Law or of magnetic monopoles.  However, one of the arguments in favor of monopoles is to be symmetric between the electric and magnetic currents. But then, if one wants to consider such symmetry one should extend it to potentials that couple to these currents. A second reason not to consider an axial potential is that in QED the chiral symmetry is broken since fermions have mass. Since it is now understood that mass can be induced by dynamics, that objection is no longer valid. Moreover, not only can mass be induced dynamically, in a conformal invariant theory mass must be induced dynamically since there can be no mass scales at the level of the Lagrangian. Since the tachyonic mass term associated with a fundamental Higgs scalar field would violate the conformal symmetry, there should be no such tachyonic term present in the Lagrangian, with all mass scales having to come from quantum fluctuations.  Finally, if one does want symmetry between the electric and magnetic sectors, with $S_{\mu}$ being able to have a geometric origin, it is thus natural to seek a geometric origin for $A_{\mu}$ too. In fact not only is it natural, that is what led Weyl to Weyl geometry in first place.

\section{Metrication of Electromagnetism}

\subsection{Implementing Conformal Invariance}

In developing Weyl geometry Weyl generalized the Levi-Civita connection  by augmenting it with  the Weyl connection to give a full connection of the form
\begin{eqnarray}
\tilde{\Gamma}^{\lambda}_{\phantom{\alpha}\mu\nu}&=&\Lambda^{\lambda}_{\phantom{\alpha}\mu\nu}+W^{\lambda}_{\phantom{\alpha}\mu\nu}
\nonumber\\
&=&\frac{1}{2}g^{\lambda\alpha}(\partial_{\mu}g_{\nu\alpha} +\partial_{\nu}g_{\mu\alpha}-\partial_{\alpha}g_{\nu\mu})
\nonumber\\
&-&g^{\lambda\alpha}(g_{\nu\alpha}A_{\mu} +g_{\mu\alpha}A_{\nu}-g_{\nu\mu}A_{\alpha}).
\label{44}
\end{eqnarray}
Weyl introduced this particular connection since under a local conformal transformation of the form 
\begin{eqnarray}
g_{\mu\nu}(x)\rightarrow e^{2\alpha(x)}g_{\mu\nu}(x),~~A_{\mu}(x)\rightarrow A_{\mu}(x)+\partial_{\mu}\alpha(x)
\label{45}
\end{eqnarray}
$\tilde{\Gamma}^{\lambda}_{\phantom{\alpha}\mu\nu}$ transforms into itself, to thus be locally conformal invariant. In consequence, the generalized Riemann tensor $\tilde{R}^{\lambda}_{\phantom{\rho}\mu\nu\kappa}$ built as per (\ref{7}) with this $\tilde{\Gamma}^{\lambda}_{\phantom{\alpha}\mu\nu}$ would be locally conformal invariant too \cite{footnote12}. However,  if one uses this $\tilde{\Gamma}^{\lambda}_{\phantom{\alpha}\mu\nu}$  connection the metric would obey  
\begin{eqnarray}
\tilde{\nabla}_{\sigma}g^{\mu\nu}(x)=-2g^{\mu\nu}A_{\sigma}(x),
\label{46}
\end{eqnarray}
with parallel transport then being path independent.

As well as develop Weyl geometry, Weyl made a particularly useful discovery for Riemann geometry itself. Specifically, he found a purely Riemannian,  Levi-Civita-based tensor, the Weyl conformal tensor, viz. 
\begin{eqnarray}
&&C_{\lambda\mu\nu\kappa}= R_{\lambda\mu\nu\kappa}
-\frac{1}{2}(g_{\lambda\nu}R_{\mu\kappa}-
g_{\lambda\kappa}R_{\mu\nu}
\nonumber\\
&&-g_{\mu\nu}R_{\lambda\kappa}+
g_{\mu\kappa}R_{\lambda\nu})
+\frac{1}{6}R^{\alpha}_{\phantom{\alpha}\alpha}(
g_{\lambda\nu}g_{\mu\kappa}-
g_{\lambda\kappa}g_{\mu\nu}),~~~
\label{47}
\end{eqnarray}
in which, remarkably, all derivatives of $\alpha(x)$ drop out identically under a local conformal transformation on the metric of the form  $g_{\mu\nu}(x)\rightarrow e^{2\alpha(x)}g_{\mu\nu}(x)$. The Weyl tensor thus bears the same relation to a local conformal transformation as the Maxwell tensor does to  a local gauge transformation, with the $I_{\rm W}=-\alpha_g\int d^4x (-g)^{1/2}C_{\mu\nu\sigma\tau}C^{\mu\nu\sigma\tau}$ Weyl action with dimensionless gravitational coupling constant $\alpha_{g}$ being the conformal analog of the  $\int d^4x (-g)^{1/2}F_{\mu\nu}F^{\mu\nu}$ Maxwell action. 

When written in terms of the Riemann tensor $I_{\rm W}$ takes the form 
\begin{eqnarray}
I_W&=&-\alpha_g\int d^4x (-g)^{1/2}C_{\lambda\mu\nu\kappa}
C^{\lambda\mu\nu\kappa}
\nonumber \\
&=&-\alpha_g\int d^4x (-g)^{1/2}\bigg{[}R_{\lambda\mu\nu\kappa}
R^{\lambda\mu\nu\kappa}
\nonumber \\
&-&2R_{\mu\kappa}R^{\mu\kappa}+\frac{1}{3}
(R^{\alpha}_{\phantom{\alpha}\alpha})^2\bigg{]}.
\label{48}
\end{eqnarray}
With $(-g)^{1/2}\left[R_{\lambda\mu\nu\kappa}R^{\lambda\mu\nu\kappa}-4R_{\mu\kappa}R^{\mu\kappa}+(R^{\alpha}_{\phantom{\alpha}\alpha})^2\right]$ being a total divergence (the Gauss-Bonnet theorem), the Weyl action can be written more compactly as 
\begin{equation}
I_W=-2\alpha_g\int d^4x
(-g)^{1/2}\left[R_{\mu\kappa}R^{\mu\kappa}-\frac{1}{3}
(R^{\alpha}_{\phantom{\alpha}\alpha})^2\right],~~
\label{49}
\end{equation}
 to give the form presented in (\ref{43}).

Weyl thus provides us with two specific ways to implement conformal invariance. To determine which one, if either, might be the relevant one for physics we need to determine how $A_{\mu}$ is to transform under a conformal transformation. To this end we look to the coupling of $A_{\mu}$ not to the geometry but  to fermions instead. And as noted above, without any reference to Weyl geometry, if we construct the Dirac action by coupling the fermion to the geometry in a  strictly Riemannian way while coupling the fermion to $A_{\mu}$ in a standard minimally coupled local electromagnetic gauge invariant way, the $\int d^4x(-g)^{1/2}i\bar{\psi}\gamma^{a}V^{\mu}_a(\partial_{\mu}+\Sigma_{bc}\omega^{bc}_{\mu}-iA_{\mu})\psi$ action that results  will be locally conformal invariant under $g_{\mu\nu}(x)\rightarrow e^{2\alpha(x)}g_{\mu\nu}(x)$, $V^a_{\mu}(x)\rightarrow  e^{\alpha(x)}V^a_{\mu}(x)$, $\psi(x)\rightarrow  e^{-3\alpha(x)/2}\psi(x)$ only if $A_{\mu}(x)$ undergoes no transformation at all. Hence immediately we see that if we want to metricate electromagnetism and recover this same Dirac action via a generalized connection, we must do so with an $A_{\mu}$ that has conformal weight zero, with  $g_{\mu\nu}$ and $A_{\mu}$ then respectively transforming as  
\begin{eqnarray}
g_{\mu\nu}(x)\rightarrow e^{2\alpha(x)}g_{\mu\nu}(x),\qquad A_{\mu}(x)\rightarrow A_{\mu}(x)
\label{50}
\end{eqnarray}
under a local conformal transformation. 

If we now introduce the Weyl connection, we need to ask whether it is possible to construct an action for the gravitational sector that would contain it and still be invariant under  (\ref{50}). Since on dimensional grounds such an action would have to be quadratic, the most general one possible would be the combination $\int d^4x(-g)^{1/2}[a\tilde{R}_{\lambda\mu\nu\kappa}\tilde{R}^{\lambda\mu\nu\kappa}+b\tilde{R}_{\mu\kappa}\tilde{R}^{\mu\kappa}+c(\tilde{R}^{\alpha}_{\phantom{\alpha}\alpha})^2]$ for some choice of the coefficients $a$, $b$, $c$. Now under (\ref{45}) this combination is invariant for any choice of $a$, $b$ and $c$. However, if $A_{\mu}$ is not to transform under a conformal transformation, this combination would need to be invariant order by order in $A_{\mu}$. For the zeroth order term we noted above that the needed combination is the one that appears in $I_{\rm W}$, to thus have $a=1$, $b=-2$, $c=1/3$, and thus to have $b=1$, $c=-1/3$ following the use of the Gauss-Bonnet theorem. The term that is linear in $A_{\mu}$ in the combination involves a cross term between the term that is zeroth order in $A_{\mu}$ and a first order term in $A_{\mu}$ that according to (\ref{8}) is a total divergence in the Levi-Civita-based $\nabla_{\mu}$. Recalling that the Riemann tensor obeys $\nabla_{\rho}R^{\rho\alpha\beta\gamma}=\nabla^{\beta}R^{\alpha\gamma}-\nabla^{\gamma}R^{\alpha\beta}$, up to a total divergence the net linear term for the combination is found to be  of the form $\int d^4x(-g)^{1/2}[(8a+2b)W_{\lambda\mu\nu}\nabla^{\lambda}R^{\mu\nu}$
$-(b+4c)W^{\lambda}_{\phantom{\lambda}\mu\lambda}\nabla^{\mu}R^{\alpha}_{\phantom{\alpha}\alpha}]$ \cite{footnote13}. Using the Bianchi identity and the explicit form for $W^{\lambda}_{\phantom{\lambda}\mu\nu}$ given in (\ref{9}), we can write the net linear term as  $\int d^4x(-g)^{1/2}4(b+4c)A_{\mu}\nabla^{\mu}R^{\alpha}_{\phantom{\alpha}\alpha}$. Since this term is not left invariant under (\ref{50}) (the Ricci scalar not being a conformal invariant), conformal invariance requires $b+4c$ be zero, to thus  require  $b=1$, $c=-1/4$. Consequently, there is no choice for the coefficients $a$, $b$ and $c$ for which both the zeroth and first order terms in $\int d^4x(-g)^{1/2}[a\tilde{R}_{\lambda\mu\nu\kappa}\tilde{R}^{\lambda\mu\nu\kappa}+b\tilde{R}_{\mu\kappa}\tilde{R}^{\mu\kappa}+c(\tilde{R}^{\alpha}_{\phantom{\alpha}\alpha})^2]$ could simultaneously obey (\ref{50}). Thus if we introduce the Weyl connection and wish to write down an action that obeys local conformal invariance as realized via (\ref{50}), the only choice is the $A_{\mu}$-independent, strictly Riemannian $I_W=-2\int d^4x (-g)^{1/2}\left[R_{\mu\kappa}R^{\mu\kappa}-(1/3)(R^{\alpha}_{\phantom{\alpha}\alpha})^2\right]$. Thus even in the presence of the Weyl connection, the only allowed conformal action in the metric sector is the one that is completely independent of the Weyl connection term. Thus if we are able to generate a Dirac action in which the $A_{\mu}$ term is associated with the Weyl connection in some way, the path integration over the fermions  would still have to lead to the separation between the $C_{\mu\nu\sigma\tau}C^{\mu\nu\sigma\tau}$ and  $F_{\mu\nu}F^{\mu\nu}$ terms that is exhibited in (\ref{43}).

\subsection{Complex Weyl Connection}

In order to be able to actually obtain an $\int d^4x(-g)^{1/2}i\bar{\psi}\gamma^{a}V^{\mu}_a(\partial_{\mu}+\Sigma_{bc}\omega^{bc}_{\mu}-iA_{\mu})\psi$ action in which the $A_{\mu}$ term  is generated geometrically, we recall, as noted above, that this cannot be done with the Weyl connection with its real $A_{\mu}$ as is,  since the Weyl connection drops out of the Dirac action identically. Now one would of course initially want to take $A_{\mu}$ to be real, since, first, it is to describe the electromagnetic field, and, second, $A_{\mu}$ plays the same role in $W^{\lambda}_{\phantom{\alpha}\mu\nu}$ as $\partial_{\mu}$ does in $\Lambda^{\lambda}_{\phantom{\alpha}\mu\nu}$. However, from the perspective of a complex phase invariance on the fermion field, minimal electromagnetic coupling is not of the form $\partial_{\mu}-A_{\mu}$ but of the form $\partial_{\mu}-iA_{\mu}$ instead, with $A_{\mu}$ being Hermitian and $iA_{\mu}$ being anti-Hermitian. Moreover, minimal coupling must be of this latter form since if $A_{\mu}$ is $PT$ even and $\partial_{\mu}$ is $PT$ odd, one needs the extra $i$ factor in order to to enforce $PT$ symmetry. 

Now precisely the same reasoning has to apply to the connection, since we had noted above that the connection has to be $PT$ odd (i.e. $i\tilde{\Gamma}^{\lambda}_{\phantom{\alpha}\mu\nu}$ has to be $PT$ even if $\tilde{I}_{\rm D}$ is to be $PT$ even). Thus, with  $\Lambda^{\lambda}_{\phantom{\alpha}\mu\nu}$ being $PT$ odd we would need $W^{\lambda}_{\phantom{\alpha}\mu\nu}$ to be $PT$ odd too. To achieve this with a $PT$ even and Hermitian $A_{\mu}$ we thus replace  $W^{\lambda}_{\phantom{\alpha}\mu\nu}$ by 
\begin{eqnarray}
V^{\lambda}_{\phantom{\alpha}\mu\nu}=-\frac{2}{3}ig^{\lambda\alpha}\left(g_{\nu\alpha}A_{\mu} +g_{\mu\alpha}A_{\nu}-g_{\nu\mu}A_{\alpha}\right),
\label{51}
\end{eqnarray}
with $V^{\lambda}_{\phantom{\alpha}\mu\nu}$ being $PT$ odd and anti-Hermitian. Insertion of  $V^{\lambda}_{\phantom{\alpha}\mu\nu}$ with its convenient $-2/3$ charge normalization into the $\delta{\Gamma}_{\mu\lambda\nu}-(\delta{\Gamma}_{\mu\lambda\nu})^{\dagger}$ term in (\ref{22}) is then found to lead to none other than the $A_{\mu}$-dependent  contribution to  $\tilde{J}_{\rm D}$ precisely as given and normalized in (\ref{33}), to thereby oblige $A_{\mu}$ to have conformal weight zero and not transform under the conformal group at all. Thus with $V^{\lambda}_{\phantom{\alpha}\mu\nu}$ we can indeed metricate electromagnetism in the fermion sector after all. Finally, with $\tilde{\Gamma}^{\lambda}_{\phantom{\alpha}\mu\nu}=\Lambda^{\lambda}_{\phantom{\alpha}\mu\nu}+K^{\lambda}_{\phantom{\alpha}\mu\nu}+V^{\lambda}_{\phantom{\alpha}\mu\nu}$ we can obtain the entire $\tilde{J}_{\rm D}$ by metrication. Thus while Weyl sidelined $W^{\lambda}_{\phantom{\alpha}\mu\nu}$ once his scale transformation on $A_{\mu}$ was reinterpreted as a minimal coupling phase transformation with  a factor $i$, we see that this same procedure applied in $V^{\lambda}_{\phantom{\alpha}\mu\nu}$ enables us to reinstate Weyl's metrication of electromagnetism after all.

With the connection $V^{\lambda}_{\phantom{\alpha}\mu\nu}$ not coupling in $\nabla_{\mu}A_{\nu}-\nabla_{\nu}A_{\mu}$ \cite{footnote14}, and with it acting in $\tilde{J}_{\rm D}$ just like conventional electromagnetic vector potential in the fermionic sector, in a universe consisting solely of fermions, gauge bosons and gravitons (with mass generation by fermion bilinear condensates), the only place where  $V^{\lambda}_{\phantom{\alpha}\mu\nu}$ could still be manifest would be in $\tilde{R}^{\lambda}_{\phantom{\rho}\mu\nu\kappa}$, i.e. in the gravitational equations of motion should they depend on the generalized connection. Since the only role of $V^{\lambda}_{\phantom{\alpha}\mu\nu}$ in the fermion sector is to act as a standard electromagnetic potential, parallel transport of fermions with a dynamics described by  $\tilde{J}_{\rm D}$ would be just the same as the conventional parallel transport of fermions in a standard Riemannian geometry in the presence of a background electromagnetic field (and its axial analog \cite{footnote15}). Likewise, parallel transport of gauge bosons would be the same as in standard Riemannian geometry, since $V^{\lambda}_{\phantom{\alpha}\mu\nu}$ drops out of $\tilde{\nabla}_{\mu}A_{\nu}-\tilde{\nabla}_{\nu}A_{\mu}$. The only problematic case would be parallel transport of the gravitational field itself. However, we have just seen that this not a problem either since the only allowed locally conformal invariant action is the purely Riemannian-geometry-based $\int d^4x(-g)^{1/2}C_{\lambda\mu\nu\kappa}C^{\lambda\mu\nu\kappa}$. Hence with an $A_{\mu}$ with conformal weight zero (and analogously for $S_{\mu}$) the theory is strictly Riemannian and no parallel transport path dependence problem can be encountered. Thus by making two key changes in Weyl's metrication program, namely replacing $A_{\mu}$ by $iA_{\mu}$ in the Weyl connection and by taking $A_{\mu}$ to have conformal weight zero, we are not only able to metricate electromagnetism in principle, but are actually able to obtain the precise structure that any such electromagnetic metrication must possess. We thus see a dual description of electromagnetism. We can induce it by a local phase transformation on the fermion field in a standard Riemannian background geometry or can obtain it by enlarging the connection to include the Weyl connection. There are no operative distinctions between the two cases  \cite{footnote16}, and for either one fermion path integration yields a purely Riemannian Weyl-tensor-based locally conformal invariant theory of gravity. 

\subsection{Non-Abelian Generalizations}

Even though the Weyl and contorsion connections involve fermionic electric and magnetic charge quantum numbers, the pure gravitational sector only involves the Levi-Civita connection. Consequently, the approach we have developed here can naturally be extended to the non-Abelian case, with the metric not being forced to acquire any internal quantum number. Thus, on putting the fermions into the fundamental representation of  $SU(N)\times SU(N)$ with $SU(N)$ generators $T^i$ that obey  $[T^i,T^j]=if^{ijk}T^k$, we replace $A_{\mu}$ by $gT^iA^{i}_{\mu}$ and $Q_{\alpha\beta\gamma}$ by $gT^iQ^i_{\alpha\beta\gamma}$, and thus replace $S_{\mu}$ by $gT^iS^{i}_{\mu}$ in the connections, to obtain a locally $SU(N)\times SU(N)$ invariant Dirac action of the form 
\begin{eqnarray}
\tilde{J}_{\rm D}&=&\int d^4x(-g)^{1/2}i\bar{\psi}\gamma^{a}V^{\mu}_a(\partial_{\mu}+\Sigma_{bc}\omega^{bc}_{\mu}
\nonumber\\
&&-igT^iA^i_{\mu} -ig\gamma^5T^iS^i_{\mu})\psi.
\label{52}
\end{eqnarray}
On doing the path integral on the fermions the previous effective action given in  (\ref{43}) is replaced by   \cite{tHooft2010a}
\begin{eqnarray}
I_{\rm EFF}&=&\int d^4x(-g)^{1/2}C\bigg{[}\frac{1}{20}\left[R_{\mu\nu}R^{\mu\nu}-\frac{1}{3}(R^{\alpha}_{\phantom{\alpha}\alpha})^2\right]
\nonumber\\
&+&\frac{1}{3}G_{\mu\nu}^iG^{\mu\nu}_i+\frac{1}{3}S_{\mu\nu}^iS^{\mu\nu}_i\bigg{]},
\label{53}
\end{eqnarray}
with an unmodified $R_{\mu\nu}R^{\mu\nu}-(1/3)(R^{\alpha}_{\phantom{\alpha}\alpha})^2$ term and the same log divergent constant $C$ as before, but with $\partial_{\mu}A_{\nu}-\partial_{\nu}A_{\mu}$ being replaced by $G_{\mu\nu}^i=\partial_{\mu}A^i_{\nu}-\partial_{\nu}A^i_{\mu}+gf^{ijk}A^j_{\mu}A^k_{\nu}$, and $\partial_{\mu}S_{\nu}-\partial_{\nu}S_{\mu}$ being replaced by $S_{\mu\nu}^i=\partial_{\mu}S^i_{\nu}-\partial_{\nu}S^i_{\mu}+gf^{ijk}S^j_{\mu}S^k_{\nu}$, just as one would want. In the same vein $X_{\mu\nu}$ of (\ref{35}) generalizes to
\begin{eqnarray}
X^{\mu\nu}_i&=&\nabla^{\mu}A^{\nu}_i-\nabla^{\nu}A^{\mu}_i
\nonumber\\
&-&\frac{1}{2}(-g)^{-1/2}\epsilon^{\mu\nu\sigma\tau}(\nabla_{\sigma}S^i_{\tau}-\nabla_{\tau}S^i_{\sigma}),
\label{54}
\end{eqnarray}
with (\ref{53}) then being written as
\begin{eqnarray}
I_{\rm EFF}&=&\int d^4x(-g)^{1/2}C\bigg{[}\frac{1}{20}\left[R_{\mu\nu}R^{\mu\nu}-\frac{1}{3}(R^{\alpha}_{\phantom{\alpha}\alpha})^2\right]
\nonumber\\
&+&\frac{1}{3}X_{\mu\nu}^iX^{\mu\nu}_i\bigg{]},
\label{55}
\end{eqnarray}
a very compact form.

\subsection{Final Comments}

As an action the effective  $I_{\rm EFF}$ contains all the symmetries of $\tilde{J}_{\rm D}$, both all its local ones and its $PT$ and $CPT$ symmetries. However, while $I_{\rm EFF}$ contains the Maxwell action, we note that it does not actually contain the Einstein-Hilbert action, and indeed it could not since the Einstein-Hilbert action is not locally conformal invariant. The gravitational action that $I_{\rm EFF}$ does  contain is locally conformal invariant, as it of course would have to be given the local conformal invariance of  the underlying $\tilde{J}_{\rm D}$. Thus we see that local conformal invariance is to gravity what local gauge invariance is to  electromagnetism, and the two are naturally linked to each other since photons and gravitons both propagate on the conformal invariant light cone. With a fermion generically transforming as $e^{\alpha_{\rm RE}}\psi $ under a conformal transformation and as  $e^{i\alpha_{\rm IM}}\psi$ under an electromagnetic gauge transformation, we thus unify gravitation and electromagnetism by gauging both the real and imaginary parts of the phase of the fermion.

The other unifying feature of the conformal gravity sector and the Maxwell sector actions given in $I_{\rm EFF}$ is that both sectors involve dimensionless couplings alone, so that as quantum theories both are renormalizable \cite{footnote17}. However, because a conformal gravity theory based on $ \int d^4x (-g)^{1/2}C_{\lambda\mu\nu\kappa}C^{\lambda\mu\nu\kappa} \equiv 2\int d^4x (-g)^{1/2}[R_{\mu\nu}R^{\mu\nu}-(1/3)(R^{\alpha}_{\phantom{\alpha}\alpha})^2]$ involves fourth-order derivative equations of motion, the theory had long been thought to possess negative norm states or negative energies. However, detailed examination of the quantization procedure revealed  \cite{Bender2008a,Bender2008b}, \cite{Mannheim2011a,Mannheim2012a} that the quantum Hamiltonian is not in fact Hermitian but is instead  $PT$ symmetric, and that when one uses the requisite $\langle L(t)|R(t)\rangle$ norm and $\langle \Omega_L|T(\phi(x)\phi(y))|\Omega_R\rangle$ type Green's functions there are then neither negative norm states nor negative energies. Consequently, conformal gravity is a fully consistent and unitary quantum theory of gravity. Interestingly for our purposes here, the key step needed to avoid negative energies was to recognize that the gravitational field $g_{\mu\nu}$ had to be an anti-Hermitian rather than a Hermitian field and be a $PT$ eigenstate \cite{footnote18}.  Intriguingly, to be able to go from $W^{\lambda}_{\phantom{\alpha}\mu\nu}$ to $V^{\lambda}_{\phantom{\alpha}\mu\nu}$ Weyl's electromagnetic field $A_{\mu}$ had to be reinterpreted in precisely the same way.

Moreover, not only is conformal gravity a consistent quantum gravity theory, there is even some encouraging observational support for it. Specifically, in  \cite{Mannheim2011b,Mannheim2012,OBrien2012,Mannheim2013} fits were provided to the rotation curves of 141 spiral galaxies using a universal formula provided by the conformal theory with only one free parameter per galaxy (the standard mass to light ratio of the luminous matter, a parameter that is common to all rotation curve studies). In the fits no need was found for any of the copious amounts of dark matter required of the standard Newton-Einstein gravity treatment of rotation curves. With current dark matter halo studies requiring two free parameters for the halo of each galaxy, to fit the same 141 galaxies dark matter fits require 282 more free parameters than conformal gravity, with the fitting thus currently favoring the conformal theory. 

To conclude we note that Weyl's ideas on conformal invariance and unification can still be of relevance  today, and could be much closer to conventional fundamental physics than had previously been thought to be the case.


\begin{thebibliography}{99}
%
\bibitem{Hehl1976}
F.~W.~Hehl,~P.~von~der~Heyde,~G.~D.~Kerlick,~and~J.~M.~Nester,
Rev. Mod. Phys. \textbf{48}, 393 (1976).
%
\bibitem{Shapiro2002}
I.~L.~Shapiro,
Phys.~Rept. \textbf{357}, 113 (2002).
%
\bibitem{Hammond2002}
R.~T.~Hammond, Rep.~Prog.~Phys. \textbf{65}, 599 (2002).
%
\bibitem{Scholz2011} 
E.~Scholz,~\textit{Weyl geometry in late 20th century physics},~arXiv:1111.3220 [math.HO], November, 2011.
%
\bibitem{Yang2014} C.~N.~Yang,~\textit{The conceptual origins of Maxwell's equations and gauge theory},~Physics Today, \textbf{67}, 45 (2014).
%
\bibitem{Bender2007} C.~M.~Bender,~Rep.~Prog.~Phys. {\bf 70}, 947 (2007).
%
\bibitem{Mannheim2006} 
P.~D.~Mannheim,~Prog. Part.~Nucl.~Phys. \textbf{56}, 340 (2006).
%
\bibitem{Mannheim2011a} P.~D.~Mannheim,  Gen. Rel. Gravit. \textbf{43}, 703 (2011).
%
\bibitem{Mannheim2012a} P.~D.~Mannheim, Found.~Phys. \textbf{42}, 388 (2012).
%
\bibitem{tHooft2010a}
G.~'t Hooft,~\textit{Probing the small distance structure of canonical quantum gravity using the conformal group},~arXiv:1009.0669 [gr-qc], September, 2010.
%
\bibitem{tHooft2010b}
G.~'t Hooft,~\textit{The Conformal Constraint in Canonical Quantum Gravity},~
arXiv:1011.0061 [gr-qc], October, 2010.
%
\bibitem{tHooft2011}
G.~'t Hooft,~Found.~Phys.~\textbf{41}, 1829 (2011).
%
\bibitem{tHooft2014}
G.~'t Hooft,~\textit{Local Conformal Symmetry: the Missing Symmetry Component for Space and Time},~arXiv:1410.6675 [gr-qc], October, 2014.
%
\bibitem{Fabbri2014} 
L.~Fabbri and P.~D.~Mannheim,~Phys.~Rev.~D \textbf{90}, 024042 (2014).
%
\bibitem{Mannheim2014a} 
P.~D.~Mannheim and J.~J.~Poveromo, Gen.~Rel.~Gravit. \textbf{46}, 1795 (2014).
%
\bibitem{Mannheim2014b} 
P.~D.~Mannheim,~\textit{Torsion, Magnetic Monopoles and Faraday's Law via a Variational Principle},~arXiv:1406.2265 [hep-th], June, 2014.
%
\bibitem{Hayashi1977}
K.~Hayashi,~M.~Kasuya, and T.~Shirafuji, Prog.~Theor.~Phys. \textbf{57}, 431 (1977).
%
%
\bibitem{Kibble1963}
T.~W.~B.~Kibble, Jour.~Math.~Phys. \textbf{4}, 1433 (1963).
%
\bibitem{Bender1998} C.~M.~Bender and S.~ Boettcher, 
Phys.~Rev.~Lett.~{\bf 80}, 5243 (1998).
%
\bibitem{Bender2002} C.~M.~Bender, M.~V.~Berry, and A.~Mandilara, J.~Phys.~A: Math.~Gen.~\textbf{ 35}, L467 (2002).
%
\bibitem{Bender2010} C.~M.~Bender and P.~D.~Mannheim,  Phys.~Lett.~A \textbf{374}, 1616 (2010).
%
\bibitem{Mannheim2013a} P.~D.~Mannheim, Phil. Trans. Roy. Soc. A \textbf{371}, 20120060 (2013). 
%
\bibitem{Bender2005} C.~M.~Bender, S.~F.~Brandt, J.-H.~Chen, and Q.~Wang,
Phys.~Rev.~D {\bf 71}, 025014 (2005).
%
\bibitem{Bender2008a} C.~M.~Bender and P.~D.~Mannheim,  Phys.~Rev.~Lett.~\textbf{100},
110402 (2008).
%
\bibitem{Bender2008b} C.~M.~Bender and P.~D.~Mannheim,  Phys.~Rev.~D \textbf{78},
025022 (2008).
%
\bibitem{footnote1} Since charge conjugation leaves the coordinates untouched, one could equally have said  that in the coordinate sector $CPT$ is compatible with Lorentz invariance. Further discussion of $T$ transformations in relativistic quantum theory may be found in C.~M.~Bender and P.~D.~Mannheim,  Phys.~Rev.~D \textbf{84}, 105038 (2011).
%
\bibitem{footnote2} Not only are the $\mathbf{E}$ and $\mathbf{B}$ fields reducible under real Lorentz transformations, they are reducible under complex Lorentz transformations as well, since if $\exp(iw_{\mu\nu}M^{\mu\nu})$ does not mix the $D(1,0)$ and $D(0,1)$ components with each other when $w_{\mu\nu}$ is real, it does not do so if $w_{\mu\nu}$ is complex. This is to be contrasted with representations that contain both left- and right-handed components such as $D(1/2,1/2)$, since here all four components do mix under real Lorentz transformations, and thus continue to do so under complex ones, with the $PT$ transformation that takes $x_{\mu}$ to $-x_{\mu}$ corresponding to a sequence of three complex Lorentz transformations $x^{\prime}=x\cosh \xi +t\sinh \xi $, $y^{\prime}=y\cosh \xi +t\sinh \xi $, $z^{\prime}=z\cosh \xi +t\sinh \xi $, each with a boost angle $\xi=i\pi$. Thus under a sequence of $PT$ transformations and Lorentz boosts with complex boost angle one can transform $\mathbf{E}(t,\mathbf{x})\pm i\mathbf{B}(t,\mathbf{x})$ first into $-[\mathbf{E}(-t,-\mathbf{x})\mp i\mathbf{B}(-t,-\mathbf{x})]$ and then into $-[\mathbf{E}(t,\mathbf{x})\mp i\mathbf{B}(t,\mathbf{x})]$.
%
\bibitem{footnote3} Since $P\psi(t,\mathbf{x})P^{-1}=\gamma_0\psi(t,-\mathbf{x})$, $T\psi(t,\mathbf{x})T^{-1}=i\gamma^1\gamma^3\psi(-t,\mathbf{x})$, $PT\psi(t,\mathbf{x})T^{-1}P^{-1}=-\gamma^2\gamma^5\psi(-t,-\mathbf{x})$, it follows that $PT(1\mp \gamma^5)\psi(t,\mathbf{x})T^{-1}P^{-1}=-(1\pm \gamma^5)\gamma^2\gamma^5\psi(-t,-\mathbf{x})$. 
%
\bibitem{footnote4} The scalar $\bar{\psi}\psi$ and pseudoscalar $\bar{\psi}i\gamma^5\psi$ will be associated with the fermion condensate mass generating mechanism that is to break the conformal symmetry dynamically.
%
\bibitem{footnote5} That $I_{\rm D}$ would have all these invariances is due to the fact that the action $(1/2)\int d^4x(-g)^{1/2}i\bar{\psi}\gamma^{a}\partial_a\psi +H. c.$ that we started with before we sought any local structure at all was that of a free flat space massless fermion field, viz. a field that is constrained to propagate on the light cone and thus possess its full conformal structure. However, we should note that transformations of the form $\psi(x)\rightarrow e^{-3\alpha(x)/2}\psi(x)$ in which the argument of the field does not change are initially somewhat different than an $x^{\mu}\rightarrow x^{\prime \mu}=\lambda x^{\mu}$ transformation since under the latter the argument of the field would change from $x^{\mu}$ to $x^{\prime \mu}$. To see that these two procedures are equivalent it is simplest to consider the free flat space massless scalar field action $I=\int d^4x(-\eta)^{1/2}\eta^{\mu\nu}\partial_{\mu}\phi(x)\partial_{\nu}\phi(x)$. With the scalar field having conformal weight equal to $-1$, under a global dilatation the action transforms into $I=\int d^4x(-\eta)^{1/2}\eta^{\mu\nu}\partial_{\mu}\phi(x^{\prime})\partial_{\nu}\phi(x^{\prime})\lambda^{-2}$. On changing the integration variable to $x^{\prime \mu}$ the action takes the form  $I=\int d^4x^{\prime}(-\eta)^{1/2}\eta^{\mu\nu}\lambda^{2}\partial^{\prime}_{\mu}\phi(x^{\prime})\partial^{\prime}_{\nu}\phi (x^{\prime})\lambda^{-2}$, to thus be invariant. However, if we define a new metric $g_{\mu\nu}=\lambda^2\eta_{\mu\nu}$ and a new field $\phi^{\prime}=\lambda^{-1}\phi$, we can rewrite the action as  $I=\int d^4x^{\prime}(-g)^{1/2}g^{\mu\nu}\partial^{\prime}_{\mu}\phi^{\prime}(x^{\prime})\partial^{\prime}_{\nu}\phi^{\prime}(x^{\prime})$, and can thus transfer the transformation on the coordinates to a transformation on the metric.
%
\bibitem{footnote6} We  also note that gauging can result in a reduction in symmetry even when internal symmetries  are involved. Consider for instance an action $\int d^4x \bar{\psi}_ii\gamma^{\mu}\partial_{\mu}\psi^i$ where $i$ runs from 1 to 8. As such, this action has a full global $SU(8)$ symmetry with the eight fermions being in its fundamental representation. And since in $SU(8)$ $8\otimes 8^*=63\oplus1$  one can gauge 63 $SU(8)$ currents and obtain a full local $SU(8)$ symmetry with 63 gauge bosons. However since the adjoint representation of $SU(3)$ is also 8 dimensional,  in the very same action one could put the eight fermions in the adjoint of $SU(3)$, and since $8 \otimes 8$ contains a symmetric $1\oplus 8\oplus 27$ part  and an  antisymmetric $8\oplus 10 \oplus 10^*$ part under $SU(3)$, one could instead  gauge the eight SU(3) currents to obtain a local $SU(3)$ gauge theory and only have eight gauge bosons. Thus after a very specific local gauging, the full global $SU(8)$ symmetry of the action is reduced to a local $SU(3)$.
%
\bibitem{Buchbinder1985} I.~L.~Buchbinder~and~I.~L.~Shapiro, Phys.~Lett.~B \textbf{151}, 263 (1985).
%
\bibitem{footnote7} The situation here is actually even simpler. Since the axial vector current involves no spacetime derivatives, no derivatives of $\alpha(x)$ would be generated in a local conformal transformation on it. Hence simply on global scale invariance grounds $S_{\mu}$ must have conformal weight zero.
%
\bibitem{footnote8} Extension to a local $SU(N)\times SU(N)$ via minimal coupling is straightforward with $A_{\mu}$ and $S_{\mu}$ being replaced in $\tilde{J}_{\rm D}$ by $gT^iA^i_{\mu}$ and $gT^iS^i_{\mu}$. Below we show that this same extension can be obtained via  metrication.
%
\bibitem{footnote9} If we introduce a vector potential $A_{\mu}$ according to $F^{\mu\nu}=\nabla^{\mu}A^{\nu}-\nabla^{\nu}A^{\mu}$ and vary with respect to $A_{\mu}$, we would only obtain $\nabla_{\nu}F^{\nu\mu}=J^{\mu}$ via  variation with  $(-g)^{-1/2}\epsilon^{\mu\nu\sigma\tau}\nabla_{\nu}[\nabla_{\sigma}A_{\tau}-\nabla_{\tau}A_{\sigma}]=0$ being satisfied identically on every variational path. In order to only have $(-g)^{-1/2}\epsilon^{\mu\nu\sigma\tau}\nabla_{\nu}[\nabla_{\sigma}A_{\tau}-\nabla_{\tau}A_{\sigma}]=0$ be obeyed at the stationary minimum alone, we need it to be non-zero away from the minimum. Since $(-g)^{-1/2}\epsilon^{\mu\nu\sigma\tau}\nabla_{\nu}[\nabla_{\sigma}A_{\tau}-\nabla_{\tau}A_{\sigma}]$ vanishes identically, we would need to relate the dual of $F_{\mu\nu}$ to some four vector other than $A_{\mu}$. As noted in \cite{Mannheim2014b} $S_{\mu}$ serves this purpose.
%
\bibitem{Shanmugadhasan1952}
S.~Shanmugadhasan, Can.~Jour.~Phys. \textbf{30}, 218 (1952).
%
\bibitem{Cabibbo1962}
N.~Cabibbo and E.~Ferrari, ~Nuovo~Cim. \textbf{23}, 1147 (1962).
%
\bibitem{footnote10} The utility of generating the magnetic monopole  sector via $S_{\mu}$ rather than by $A_{\mu}$ is that it does not require $A_{\mu}$ to have either the singularities (Dirac string) or the non-trivial topology (grand unified monopoles) that are used in order to evade the vanishing of $(-g)^{-1/2}\epsilon^{\mu\nu\sigma\tau}\nabla_{\nu}[\nabla_{\sigma}A_{\tau}-\nabla_{\tau}A_{\sigma}]$. In the $S_{\mu}$ case $(-g)^{-1/2}\epsilon^{\mu\nu\sigma\tau}\nabla_{\nu}[\nabla_{\sigma}A_{\tau}-\nabla_{\tau}A_{\sigma}]$ does vanish identically, with the monopole sector not being associated with $A_{\mu}$ at all. A second benefit to introducing $S_{\mu}$ is that the action in (\ref{39}) is  renormalizable. 
%
\bibitem{footnote11} Since both $A_{\mu}$ and $S_{\mu}$ couple to the fermionic currents in $\tilde{J}_{\rm D}$, through fermionic loops one could have transitions between the  $A_{\mu}$ and $S_{\mu}$ sectors.
%
\bibitem{footnote12} As noted in \cite{Fabbri2014}, if one sets $q=1$ in (\ref{32}) the spin connection associated with the connection $\tilde{\Gamma}^{\lambda}_{\phantom{\alpha}\mu\nu}=\Lambda^{\lambda}_{\phantom{\alpha}\mu\nu}+K^{\lambda}_{\phantom{\alpha}\mu\nu}$ would be locally conformal invariant too, as would then be the generalized Riemann tensor as built from this particular spin connection. 
%
\bibitem{footnote13} Our discussion here follows a similar discussion  for theories with torsion that was given in \cite{Fabbri2014}. 
%
\bibitem{footnote14} It is actually unnecessary to show that the Weyl connection decouples from $F_{\mu\nu}$, since in generalizing beyond standard Riemanian geometry one can only replace the Levi-Civita connection by a generalized connection in those places where the Levi-Civita connection actually appears. Since the Levi-Civita connection decouples from $F_{\mu\nu}$ in a standard Riemannian geometry where $\nabla_{\mu}A_{\nu}-\nabla_{\nu}A_{\mu}=\partial_{\mu}A_{\nu}-\partial_{\nu}A_{\mu}$, there is no  Levi-Civita connection to generalize. While this does  not matter for the Weyl connection since it would decouple anyway because of its symmetry, it does matter for the torsion connection since its antisymmetry structure would permit it to couple, with $\tilde{\nabla}_{\mu}A_{\nu}-\tilde{\nabla}_{\nu}A_{\mu}$ then being given by $\partial_{\mu}A_{\nu}-\partial_{\nu}A_{\mu} +Q^{\lambda}_{\phantom{\lambda}\mu\nu}A_{\lambda}$. However, this is not the correct definition of $F_{\mu\nu}$ in the torsion case, and indeed it could not be since it would not be gauge invariant, so even in the torsion case one has to set $F_{\mu\nu}=\partial_{\mu}A_{\nu}-\partial_{\nu}A_{\mu}$. Moreover, if one then takes the action to be of the form $-(1/4)\int d^4x(-g)^{1/2}(\partial_{\mu}A_{\nu}-\partial_{\nu}A_{\mu})(\partial^{\mu}A^{\nu}-\partial^{\nu}A^{\mu})$, the Maxwell equations that are then produced by variation with respect to $A_{\mu}$ will only depend on the Levi-Civita connection derivative and be of the form $\nabla_{\nu}(\partial^{\nu}A^{\mu}-\partial^{\mu}A^{\nu})=\partial_{\nu}(\partial^{\nu}A^{\mu}-\partial^{\mu}A^{\nu})+
(-g)^{-1/2}\partial_{\nu}(-g)^{1/2}(\partial^{\nu}A^{\mu}-\partial^{\mu}A^{\nu})=0$, to thus be independent of the generalized connection altogether.
%
\bibitem{footnote15} While it is intriguing to give electromagnetism such a chiral structure, we need to explain  why there is no sign of any axial massless photon, and why its presence would not impair the great success achieved by a quantum electrodynamics  that is based purely on $A_{\mu}$ alone. To this end  it was suggested in \cite{Mannheim2014b} that the axial symmetry is spontaneously broken with $S_{\mu}$ acquiring a Higgs mechanism type mass. On noting that $\bar{\psi}\gamma^{\mu}\psi A_{\mu}+\bar{\psi}\gamma^{\mu}\gamma^5\psi S_{\mu}=
(1/2)\bar{\psi}(\gamma^{\mu}-\gamma^{\mu}\gamma^5)\psi(A_{\mu}-S_{\mu})+(1/2)\bar{\psi}(\gamma^{\mu}+\gamma^{\mu}\gamma^5)\psi(A_{\mu}+S_{\mu})$, we see that a straightforward way to implement a Higgs mechanism for $S_{\mu}$ is to embed not just $A_{\mu}$ but also $S_{\mu}$ into a non-Abelian chiral weak interaction such as the $SU(2)_{\rm L}\times SU(2)_{\rm R}\times U(1)$ type theory discussed in P.~D.~Mannheim, Phys.~Rev.~D \textbf{22}, 1729 (1980) and references therein.  An advantage of doing this is that if the theory is broken down to $SU(2)_{\rm L}\times U(1)$ by making right-handed gauge bosons very heavy, this would explain the lack of detection to date of the right-handed neutrinos that are required by the conformal symmetry. (If the chiral symmetry breaking is achieved by giving a right-handed  neutrino Majorana mass $\psi^{\rm Tr}(1+\gamma^5)i\gamma^2\gamma^0(1+\gamma_5)\psi$ a non-zero vacuum expectation value, then since its $PT$ transform is  $\psi^{\rm Tr}(1-\gamma^5)i\gamma^2\gamma^0(1-\gamma_5)\psi$, $PT$ would be spontaneously broken too.) Thus, rather than being some arcane geometrical curiosity, because of its association with a metrication of $S_{\mu}$, torsion would actually be manifest as a perfectly normal and even quite mundane gauge boson that gets its mass via the Higgs mechanism. Thus if we seek a metrication of the fundamental forces through the Weyl and torsion connections, we are led to a quite far reaching conclusion, namely that not only must the fundamental forces be described by local gauge theories, they must be described by spontaneously broken ones. (It was also suggested in \cite{Mannheim2014b}  that intrinsically antisymmetric  torsion might instead have escaped detection by being based on hard to detect anticommuting Grassmann numbers.)

%
\bibitem{footnote16} From the perspective of minimal coupling there is however a distinction in principle, since one is not actually obliged to couple electromagnetism minimally at all as one could introduce a fundamental $(e/m)\bar{\psi}F^{\mu\nu}i[\gamma_{\mu},\gamma_{\nu}]\psi$ type coupling into electromagnetism as well. However such a coupling is not generated geometrically via the Weyl connection, and could anyway not be generated in a conformal invariant theory since given its $m$ dependence,  the coupling is not conformal invariant.
%
\bibitem{footnote17} While the effective $I_{\rm EFF}$ action given in  (\ref{53}) and (\ref{55}) is motivated by local conformal invariance and the generalized Weyl and torsion connections, we note that it is actually more general than that. Specifically, while this $I_{\rm EFF}$ arises as the one fermion loop radiative correction to the $\tilde{J}_{\rm D}$ action given in (\ref{52}), actions such as the $\tilde{J}_{\rm D}$ action itself will arise in any local non-Abelian gauge theory even if the connection is just the Levi-Civita one. In other words this action  is not just a standard action, but with the appropriate non-Abelian gauge group, it is the one that is expressly used for the  fundamental forces. Hence, regardless of what explicit form the gravitational sector action might take, the $I_{\rm EFF}$ action given in (\ref{53}) will always be generated in any gravitational theory. Thus no matter what the gravity theory, one will always have to deal  with a log divergent radiatively-induced conformal gravity action. Moreover, as noted in \cite{tHooft2010a} radiative loops due to other standard fields such as scalars and gauge bosons yield a log divergence of the same sign, and thus the fermionically generated $I_{\rm EFF}$ could not be cancelled by other fundamental fields. To cancel this divergence one must therefore introduce a counter term of exactly the same form as $I_{\rm EFF}$, and thus one must introduce the  $I_{\rm W}$ Weyl action given in (\ref{49}) into the theory. If that is all that one introduces, one then has a fully renormalizable quantum gravitational theory.
%
\bibitem{footnote18} If  we replace $g_{\mu\nu}$ by $ig_{\mu\nu}$, and thus $g^{\mu\nu}$ by $-ig^{\mu\nu}$ (since $g^{\mu\lambda}g_{\lambda\nu}=\delta^{\mu}_{\nu}$), then neither the connection nor the Riemann tensor undergo any change. Standard gravitational measurements are thus insensitive as to whether the overall phase of the gravitational field is real or purely imaginary, with the phase only being measurable via interference with another field such as the electromagnetic one.
%
\bibitem{Mannheim2011b}
P.~D.~Mannheim~and~J.~G.~O'Brien, Phys.~Rev.~Lett. \textbf{106}, 121101 (2011).
%
\bibitem{Mannheim2012}
P.~D.~Mannheim~and~J.~G.~O'Brien,~Phys.~Rev.~D \textbf{85}, 124020 (2012).
%
\bibitem{OBrien2012}
J.~G.~O'Brien~and P.~D.~Mannheim, Mon.~Not.~R.~Astron.~Soc. \textbf{421}, 1273 (2012).
%
\bibitem{Mannheim2013}
P.~D.~Mannheim~and~J.~G.~O'Brien,~J.~Phys.~Conf.~Ser. \textbf{437}, 012002 (2013). 
%
\end{thebibliography}
\end{document}